\documentclass[12pt,preprint]{aastex}
\usepackage{psfig}
\usepackage{longtable}

\def\hb{\ifmmode {\rm H}\beta \else H$\beta$\fi}

\def \L5100{$L_{5100}$}

\def \Ledd{$L/L_{\rm Edd}$}

\def  \kms         {\hbox{km s$^{-1}$}}

\def \spitzer      {{\it Spitzer}}

\slugcomment{}

\shorttitle{AGN/Star formation connection in NLS1} 
\shortauthors{E. Sani et al.}

\begin{document}

\title{Enhanced Star Formation in Narrow Line Seyfert 1 AGN revealed by Spitzer}

\author{E. Sani\altaffilmark{1,2}, D. Lutz\altaffilmark{2}, G. Risaliti,\altaffilmark{3,4}, H. Netzer\altaffilmark{5}, L. C. Gallo\altaffilmark{6}, B. Trakhtenbrot\altaffilmark{5}, E. Sturm\altaffilmark{2}, T. Boller\altaffilmark{2}}
%\\ 
%2pm 21 May 2003}

\email{sani@arcetri.astro.it}

\altaffiltext{1}{Dipartimento di Astronomia, Universit\`a di Firenze, Largo
E. Fermi 2, I-50125 Firenze, Italy}
\altaffiltext{2}{Max-Planck-Institut f\"{u}r extraterrestrische Physik, Postfach 1312, 85741 Garching, Germany}
\altaffiltext{3}{INAF - Osservatorio Astrofisico di Arcetri, Largo
E. Fermi 5, I-50125 Firenze, Italy}
\altaffiltext{4}{Harvard-Smithsonian Center for Astrophysics, 60
Garden Street, Cambridge, MA 02138}
\altaffiltext{5}{School of Physics and Astronomy and the Wise Observatory, The Raymond and Beverly Sackler Faculty of Exact Sciences, Tel-Aviv University, Tel-Aviv 69978, Israel}
\altaffiltext{6}{Department of Astronomy \& Physics, Saint Mary's University, 923 Robie Street, Halifax, NS B3H 3C3, Canada}

\begin{abstract}
We present new low resolution \spitzer\ mid-infrared spectroscopy of a sample 
of 20 ROSAT selected local Narrow Line Seyfert 1 
galaxies (NLS1s). We detect strong AGN continuum in all and clear PAH 
emission in 70\% of the sources.
The 6.2 $\mu$m PAH luminosity spans three orders of magnitudes, from 
$\sim10^{39}$erg~s$^{-1}$ to $\sim10^{42}$erg~s$^{-1}$ 
providing strong evidence for intense ongoing star formation in the 
circumnuclear regions of these sources.\\
Using the IRS/\textit{Spitzer} 
archive we gather a large number of additional NLS1s and 
their broad line counterparts (BLS1s) and constructed NLS1 and BLS1 
sub-samples to compare them in various ways. The comparison shows a 
clear separation according to FWHM(\hb) such that objects with 
narrower broad \hb\ lines are the strongest PAH emitters.
We test this division in various ways trying to remove biases due to 
luminosity and aperture size. Specifically, we find that star formation 
activity around NLS1 AGN is larger than around BLS1 of the same AGN luminosity.
The above result seems to hold over the entire range of distance and 
luminosity. 
Moreover the
star formation rate is higher in low black hole mass and high \Ledd\
systems indicating that black hole growth and star formation are occurring
simultaneously.
\end{abstract}
%%%%%%%%%%%%%%%%%%%%%%%%%%%%%%%%%%%%%%%%%%%%%%%%%%%%%%%%%%%%%%%%


%%%%%%%%%%%%%%%%%%%%%%%%%%%%%%%%%%%%%%%%%%%%%%%%%%%%%%%%%%%%%%%%

\section{Introduction}

The connection and co-evolution of active galactic nuclei (AGN) and 
star forming or starburst (SB) galaxies has 
become a major area of research. Understanding this evolution will supply 
most valuable information about the formation of the first black holes (BH), 
the accretion rate history and the growth of such systems
through time, the building of galactic bulges, the metal enrichment of 
galaxies and clusters and more.
Early studies (e.g. Rowan-Robinson 1995) suggested a possible common fueling 
mechanism for starbust galaxies and quasars. 
There is mounting evidence that intense star formation and nuclear activity 
are often closely related. This is evident from direct optical and infrared 
(IR) observations of many AGNs, the tight relationship between 
central BH mass and bulge mass (or bulge stellar velocity dispersion;
see Ferrarese $\&$ Merrit 2000, Gebhardt et al. 2000, Marconi $\&$ Hunt 2003)
and also from the analysis of luminous sources such as ULIRGs where nuclear 
activity and intense star formation (SF) are happening at the same time (e.g., Genzel et al.~1998).

The long term interest in the properties of type-I AGNs with narrow broad 
emission lines (Narrow Line Seyfert 1 galaxies - NLS1s) is closely related 
to the above issues. Such systems were defined more than three 
decades ago (Zwicky et al.~1971) and have received much attention 
due to their unusual X-ray properties. They are normally defined by their 
optical spectra in particular the requirement that FWHM(\hb)$\leq 2000$ \kms (Osterbrock $\&$ Pogge 1985).
Some properties of NLS1 are definitely extreme among type-I AGNs:
(I) their X-ray spectra are very steep showing the presence of a large 
soft X-ray excess (Boller et al. 1996). The X-ray slope defined by 
$ N_{ph} \propto E^{-\Gamma}$, has photon spectral indices ranging from 
$\Gamma=1$ to $\Gamma>4$. The X-ray luminosity span the range 
$10^{42}-10^{45}$~erg~s$^{-1}$.
(II). They exhibit rapid, large amplitude X-ray variability 
(Gallo et al. 2004 and references therein). 
(III). Their optical FeII lines are very strong compared with \hb (Grupe et al. 2004).
(IV). They show indications for large metal abundance in optical/IR
(Shemmer $\&$ Netzer 2002, Shemmer et al. 2004) and X-ray (Tanaka et al.~2004, Fabian et al.~2009). 
Other extreme properties have been noted too. For example, 
Peterson and collaborators (2000) proposed that NLS1s have lower black hole (BH) masses 
compared with broad line Seyfert 1 galaxies (hereafter BLS1s) of similar 
luminosity. Translating their accretion rate and BH mass to \Ledd\ one 
finds that they are at the high end of the AGN distribution showing 
\Ledd$\simeq 1$ (see also Boroson 2002; Grupe 2004 and references therein). 
As such, they have been considered in several papers to be in the phase 
of building up their central BH (Mathur et al. 2001; Grupe $\&$ Mathur 2004).

The large accretion rate and high metallicity of NLS1s are likely related 
to the presence of active star forming regions in the hosts of such 
sources and Mathur (2000) suggested that NLS1s live in rejuvenated,
gas-rich galaxies with ongoing star-formation. 
This has been speculated in several individual sources with an ULIRG 
infrared nature (e.g. Tacconi
et al.~2002; Risaliti et al. 2006) but has not been shown to be the case 
in large NLS1s samples. 

In the present work, we want to obtain a detailed mid-infrared (MIR) 
spectroscopic characterization of circumnuclear regions in the hosts of NLS1s.
 The aims are to investigate the possibility that such hosts show more 
intense star formation compared with other AGNs
 and to check the various relationships between SF and BH accretion. 
The high sensitivity and resolution of the IRS spectrometer on board the 
\textit{Spitzer Space Telescope} (Houck et al. 2004) provide a most suitable 
tool to achieve these goals by obtaining high S/N low resolution 5-15~$\mu$m 
spectra of many NLS1s. In particular we want to characterize the SF in such 
system using observations of polycyclic aromatic hydrocarbons (PAH) features 
that are superimposed on the strong MIR AGN continuum. The analysis of the 
SF-BH activity connection can then be carried out using a comparison
sample of BLS1s of comparable luminosity.

This paper is organized as follows.
The second section describes the NLS1 and BLS1 samples including the 
new {\it Spitzer} observations.
In \S3 we describe the data reduction and the analysis of the spectra
which are a superposition of the spectra of the AGN and the star forming host.
In \S4 we use our samples to investigate in detail the differences 
between NLS1s and BLS1s and in \S5
we summarize our results and list the most important conclusions.
Throughout this paper we assume $H_0=70$~km~s$^{-1}$~Mpc$^{-1}$, $\Omega_M=0.3$, $\Omega_\Lambda=0.7$.

\section{The sample selection}

Our prime sample of 20 NLS1s is taken from a ROSAT sample of nearby 
AGN (Thomas et al.~1998), carefully characterized both in the optical 
(Grupe et al.~2004) and X-ray bands (Boller et al.~1996, Grupe et al.~2004) 
and not previously observed with \textit{Spitzer} IRS.
%Of these, 6 galaxies have already been observed by \textit{Spitzer}/IRS 
%with sufficient S/N and are part of the complete NLS1 sample
%(see \S2.1 for details). 
%The sample included six sources (I~Zw~1, PG~1402+261, PKS~0558-504, IRAS 13348+2438, ..., ...) 
%that by early 2005 had already been 
%scheduled for sufficiently deep observations by \textit{Spitzer}/IRS 
%and are part of the complete NLS1 sample (see below for details). 
The \textit{Spitzer} IRS observations of the new 20 NLS1s completing the
sample are part of the PID 20241 program (PI D. Lutz).

All 20 sources considered here are of low redshift and only one (PHL~1092) 
has $z>0.1$). The FeII line strengths range from FeII/\hb =0.1 to 
FeII/\hb =3, the soft X-ray slope is in the range $1<\Gamma<4$, and 
BH masses and X-ray luminosities span more than three order of magnitudes. 
These are some of the best studied NLS1s, permitting us to compare our 
MIR observations with other multi-wavelength properties of the sources.
All our targets are relatively bright in the far infrared (FIR) and are 
detected by the IRAS satellite with a 60~$\mu$m flux 0.2~Jy$<$S$_{60}<$7~Jy. 
About 2/3 are detected at 12~$\mu$m with flux greater than 0.1~Jy.

The new data described here were supplemented by a large number of 
additional {\it Spitzer} archival spectroscopic observations of NLS1s and 
BLS1s. These were selected as follows:
We first extracted all the NLS1s and BLS1s from the $12^{th}$ edition of 
the Catalog of Quasars and Active Nuclei compiled by V{\'e}ron-Cetty \& 
V{\'e}ron (2006, hereafter VQC06), using both catalogs of faint (i.e. Seyfert 
galaxies fainter than absolute magnitude M$~=~23$) and bright AGNs (i.e. 
quasars brighter than M$~=~23$). We probed for `S1n' and for `S1' or `S1.0' or 
`S1.2' classifications in VQC06, respectively. The defining criterion for 
inclusion as 
an NLS1 in the VQC06 is FWHM(\hb)$\leq 2000$ \kms. To assure high quality 
spectra, and to reduce aperture effects (see below) we used an upper
redshift limit of z$~\leq~0.2$. The final samples cover six orders of 
magnitude in luminosity. Among the BLS1s listed in VQC06, we excluded 
sources with Seyfert subclasses 1.5, 1.8 and 1.9. This is done to avoid
confusion between NLS1s and BLS1s with particularly strong narrow lines, 
and possible intrinsic differences between Type 1 and Type 2 objects. We verified VQC06
spectral classifications from the literature or from SDSS DR7 spectra, 
eliminating a number of objects classified as S1 (1.0 or 1.2) in VQC06 but in fact 
corresponding to intermediate types lacking detectable BLR emission in \hb. 

Radio loudness is another potential bias. Synchrotron emission in radio 
loud (RL) AGNs can extend to shorter wavelengths and significantly
contribute to the observed MIR continuum. This will dilute the hot-dust 
produced continuum and affect the relative strength of the observed PAH 
features. We therefore exclude RL sources from our sample.

The two lists thus defined were cross-checked against the Science Data 
Archive of the \textit{Spitzer} Science Center (SSC) using the 
\textit{Leopard} Archive Tool version 18.2 for selecting objects observed 
in the IRS Low Resolution Mode.\\ 
Table~\ref{tab1} shows the final samples together 
with their relevant optical properties and the observing log is in Table~\ref{tab2}.\\
While cross-checking the catalogs and archives is generally very successful, 
there are definitely some exceptions and biases related to the way the 
catalog and archive proposals were produced. This makes it necessary to 
check the selection by hand. We checked a posteriori the optical classification 
of our sources considering FWHM(\hb). 
The \hb\ measurements are from various sources with literature references 
listed in Table~\ref{tab1} and
our own measurements based on SDSS DR7 spectra and PG Quasar 
spectra (Boroson \& Green~1992, and T. Boroson priv. comm.). 
%The MIR luminosities are computed as described in \S3.
\begin{figure*}[!h]
\begin{center}
% \vspace{-.5cm}
\includegraphics[scale=0.50]{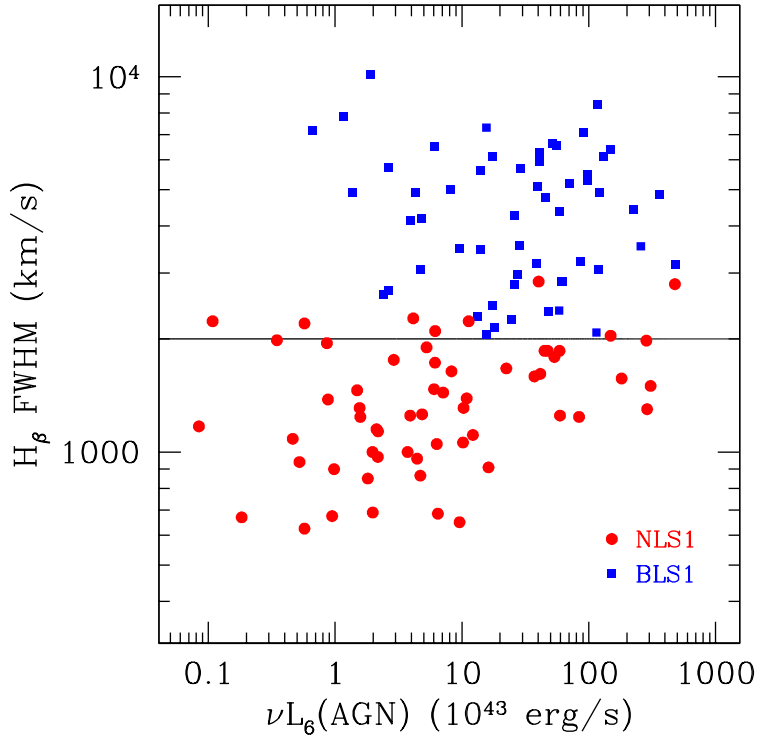}
\end{center}
% \linespread{1.0}
% \vspace{-.8cm}
\caption{FWHM(\hb) - L(6~$\mu$m) plot for our sample. 
The observed 6~$\mu$m luminosities are computed as described in \S3.
BLS1s are marked with filled blue squares and NLS1s with red symbols.
 The threshold of 2000 \kms\ is also marked.} 
% \linespread{1.6}
\label{fig1}
\end{figure*}
As shown in Fig.~\ref{fig1} and Tab.~\ref{tab1} and \ref{tab2}, 
the sub-classification into the two 
groups is not precise. Indeed 
the NLS1s Mrk~291, Mrk~705, SDSS J131305.80+012755.9, 
IRAS~13349+2438, Mrk~734, PG~1402+261, 
PG~2130+099 show FWHM(\hb) which exceeds our limit of 2000 \kms. These 
sources show most of the well known extreme NLS1 properties
such as a steep soft X-ray slope and strong FeII lines.
Indeed, all those sources are included in well known NLS1s samples such as 
 V{\'e}ron-Cetty et al.~2001, Boller et al.~1996, Zheng et al.~2002, Zhou et al.~2006 and 
Gallo et al.~2006. 
On the other hand, only sources with strictly FWHM(\hb)$> 2000$ \kms are considered 
to build the broad line comparison sample.
We note that several sources classified in the VCV06 catalog as type 1 and 
confirmed to have BLR \hb\ emission still 
show FWHM(\hb)$\leq 2000$ \kms. To be consistent, we 
move to the narrow line sample all the sources showing typical 
optical NLS1 signatures (i.e. spectra with FWHM(\hb)$\leq 2000$ \kms and intense iron emission) 
also supported by published analysis (see Table~\ref{tab1} for detailed references). 
Examples of such cases are SDSS~J131305.80+012755.9, SDSS~J142748.28+050222.0, PG~1552+085.

\section{Data reduction and analysis}
\begin{figure*}[!h]
\begin{center}
% \vspace{-.5cm}
\includegraphics[scale=0.50]{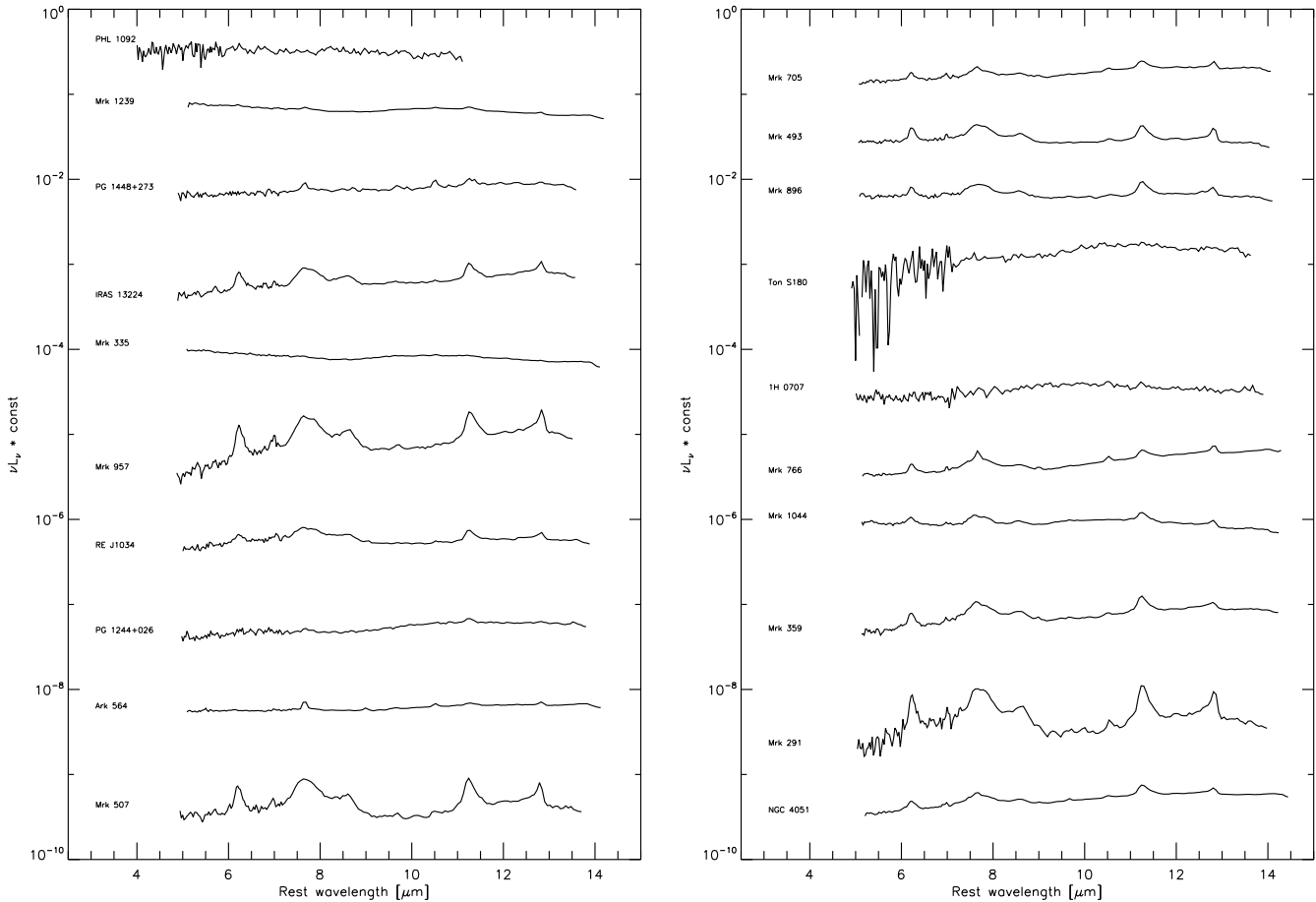}
\end{center}
% \linespread{1.0}
% \vspace{-.8cm}
\caption{\spitzer\ IRS spectra for our prime NLS1 sample. 
The spectra are arbitrarily scaled for clarity, and are ordered with
decreasing luminosity from the top left. 
These spectra are of high quality and high S/N, and are representative of the variety of NLS1 mid-IR spectral shapes.}
% \linespread{1.6}
\label{fig2}
\end{figure*}

The reduction and analysis described in this section have been
applied to our new observations as well as to the archival data. This is 
done in order to ensure a consistent and uniform analysis. All data are 
low resolution ($\lambda/\Delta\lambda\sim60-120$, long slit \textit{Spitzer} 
IRS observations obtained in a staring mode (Houck et al.~2004)). 
We are interested in the rest frame wavelength range 5-14~$\mu$m, 
corresponding to the Short Low (SL) channel. The data reduction is carried 
out by applying our own IDL-based procedures starting with the 
two-dimensional basic calibrated data (BCD) provided by the version 17 of 
the \textit{Spitzer} pipeline. We subtracted the sky by differencing, for 
each cycle, the two nod positions. In the difference thus computed, we 
replaced deviant pixels with values representative of their neighborhoods 
in the dispersion directions. In averaging all the cycles of the 
two-dimensional subtracted frames, we excluded values more than 2.5 times 
the local noise away from the mean. The calibrated one-dimensional spectra 
for the positive and negative beam were extracted using SMART (Higdon et 
al. 2004), and the two spectra averaged in order to obtain the final 
spectrum. Small multiplicative corrections where applied to take into 
account flux offset between the two low resolution orders. We also checked 
for consistency with the 12~$\mu$m and 25~$\mu$m IRAS fluxes when 
available, considering the possibility of aperture effects for nearby sources.
The observation of Ton~S180 represents a particular case. For the 
observation of this source the slit was not properly centered on the target
due to adoption of an incorrect literature position. To properly calibrate 
the flux of this source we used the Long Low channel data (rest wavelength 
range 14-35~$\mu$m) and re-normalized the observed flux to the IRAS values. 
For this reason, we consider the Ton~S180 measurements to be less reliable than the rest of the sample.
Fig.~\ref{fig2} shows the final spectra of our 20 prime sources.
We decomposed the AGN and star formation contribution to the observed 
spectrum by fitting the 5.5--6.85$\mu$m rest wavelength range of the
spectra with a simple model of a linear continuum representing thermal 
emission by AGN heated dust and a starburst template obtained from the
ISO/SWS spectrum of M82. The latter is designed to enable a reliable 
measure of the 6.2$\mu$m PAH emission feature in the spectrum. We note 
that the broad wavelength MIR spectrum of luminous AGNs (e.g. Schweitzer 
et al. 2006 (hereafter S06); Netzer et al.~2007) is well represented by a linear or 
power-law continuum over the limited wavelength range considered here.

The fit parameters obtained in this way were used to calculate the rest 
frame 6~$\mu$m pure AGN flux density and the 6.2~$\mu$m PAH flux. The latter is 
obtained by integrating over the rest wavelength range 6.1-6.35~$\mu$m. The 
detection criteria for both the AGN continuum and the 6.2~$\mu$m PAH feature
are fixed at a 3~$\sigma$ level, and the error estimates are based on propagating 
the noise estimated from the difference of spectrum and fit over 
the rest wavelength range 5.5-6.85~$\mu$m. 
Cases of non-detections were assigned a 
3~$\sigma$ continuum and/or PAH upper limit. All fluxes were converted to 
luminosities which are referred to later on as L(6$\mu$m) and L(6.2~PAH). 
The only galaxy with a continuum upper limit is considered a non detection 
and is reported in the tables for completeness but is not considered 
in the analysis and discussion sections.
The choice of the 6$\mu$m continuum enables to define in a clean way the 
hot-dust continuum produced by the dusty structure (torus) surrounding 
the AGN. This part of the spectrum is easy to measure and not much affected
by strong emission or absorption features (e.g. S06). As for the PAH, 
we prefer to use the 6.2~$\mu$m PAH feature, instead of the strong 
7.7~$\mu$m PAH emission feature, because it is not affected by blending 
with atomic lines such as [NeVI] and because its profile is not distorted 
by the 9.7~$\mu$m silicate feature. The choice of 6.2~$\mu$m PAH 
causes nondetection in some weak PAH sources where other PAH features are seen, 
we consider this acceptable and justified by its other advantages. 
The continuum and PAH fluxes (luminosities) and 
upper limits are listed in Table~3 for the complete ROSAT and archival
sample of 59 NLS1 (58 detections) and 54 BLS1s (all detected).
\section{Results}
The results of our analysis are shown in Table~3, Figures~\ref{fig3}, \ref{fig4}, 
\ref{fig6} and are summarized below:\\
- For the entire samples, the PAH features for star formation are detected with a $3~\sigma$ 
significance in 70~$\%$ of NLS1s and 45~$\%$ of BLS1s.\\
- The comparison between unbiased NLS1 and BLS1 samples shows how the two population are 
separated and that NLS1s are related to a more intense SB activity than BLS1s.\\
Possible luminosity and distance effects are carefully taken into account in the following discussion.

\subsection{Mid-IR spectral properties}
Our original sample of 20 sources is characterized by high quality and high S/N spectra, 
and can be considered representative of MIR spectral properties of the entire NLS1 sample.
As one can infer from Fig.~\ref{fig2}, and the dedicated part of 
Table~3, thanks to the high quality of our 20 spectra we are 
able to detect the 6.2~$\mu$m PAH feature in most of all the sources (14/20 of the prime sample).\\
From a visual inspection of Fig.~\ref{fig2}, 
IRAS~13224-3809, Mrk~957, Mrk~507 and Mrk~291 show the most prominent PAH features. 
Indeed, among the known AGN these belong to the ones with the 
strongest SF signatures relative to the underling AGN continuum 
(values in column 4 and 5 of Tab.~3, see below for details on R values).\\
There are only six (30\% of the subsample) upper limits for the 6.2 PAH flux. 
Among these, three spectra (PHL~1092, Ton~S180, 1H~0707-495) have a low S/N due 
to their large distance or intrinsic faintness and, in the case of Ton~S180 pointing problems (see \S3 for details), 
thus their upper limits are not very constraining (see the high values in Tab.~3 column 3).
The other limits (Ark~564, Mrk~335, PG~1448+273) have high S/N spectra and the 
low limits values establish strong constraints for the detection of the 6.2 PAH. 
Nevertheless, the lack of this feature does not imply the complete absence of ongoing star formation 
as the it can be diluted by the strong AGN continuum (S06). 
The presence of 7.7, 11.3 and 12.7 PAH features strengthens this possibility, 
Mrk~335 being the only exception. 
The peculiar case of Mrk~335 is discussed in Appendix togher with 
two sources showing strong 9.7~$\mu$m silicate absorption and one strong star forming BLS1. 

\subsection{Star Formation activity vs. FWHM(\hb) for NLS1 and BLS1 galaxies}
\begin{figure*}[!h]
\includegraphics[scale=0.45]{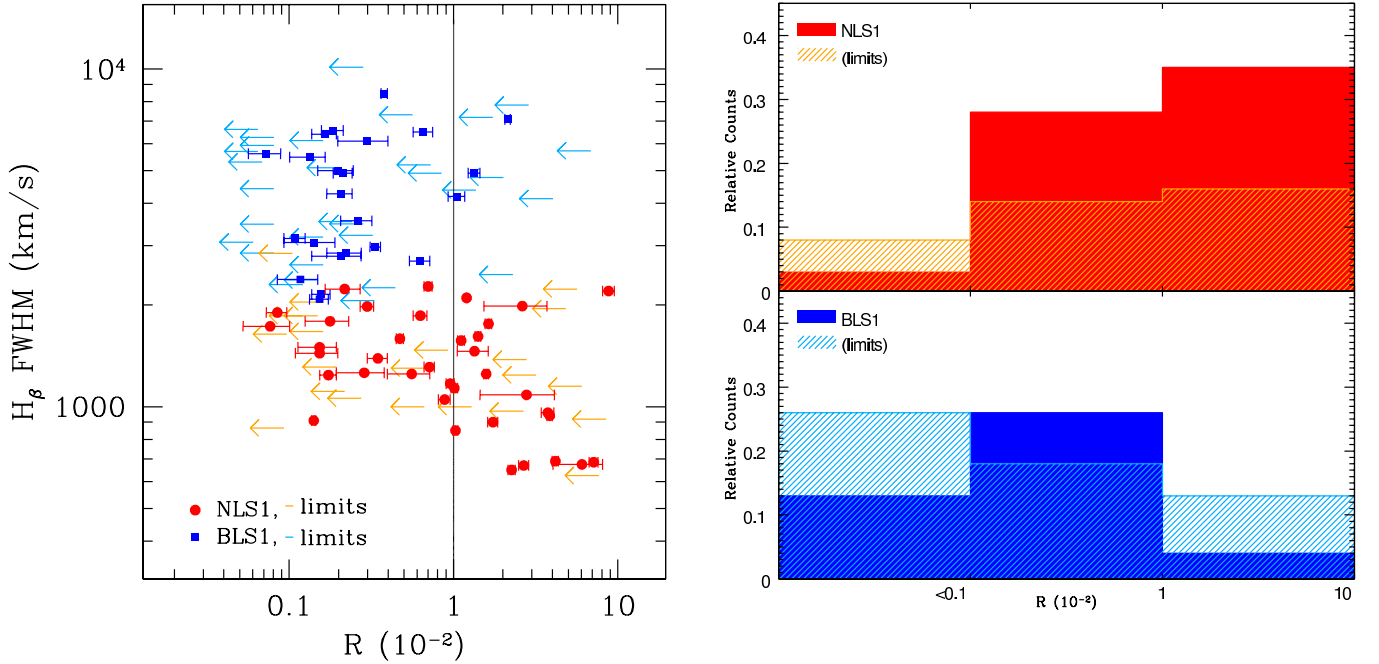}
\caption{FWHM(\hb) vs the PAH to continuum ratio R (defined in \S4.2, Eq.~\ref{eq1}). 
Arrows indicate upper limits. Red dots are NLS1 whose upper limits 
are in orange. Blue squares are BLS1 with sky-blue limits. \textit{Left panel}:
the bidimensional plot displays the measurements and the black vertical line divides 
sources with strong SF (R$>10^{-2}$) from the rest of the sample. \textit{Right panel}: 
the R relative counts (limits) distribution is shown for both the NLS1 and BLS1 samples. 
The distribution is shown in bins for the three R 
orders of magnitude, relative to the proper sample number of sources.} 
% \linespread{1.6}
%\label{fg:h_pah}
\label{fig3}
\end{figure*}

To quantify the contribution of star forming regions to the NLS1s and 
BLS1s spectra we define the SF-AGN ratio R by
\begin{equation}
 R=\frac{L (6.2,PAH)}{\nu L_\nu (6,AGN)}
\label{eq1}
\end{equation}
R is proportional to the PAH equivalent width (EW) measured against the pure 
AGN continuum. R values for the entire sample are listed in column 6 of Tab.~3, 
the correlation of R vs. FWHM(\hb) is plotted in the left panel of Fig.~\ref{fig3}, 
while histograms in the 
right panel show the distribution of R for NLS1 and BLS1.

Visual inspection of Fig.~\ref{fig3} suggests that NLS1s are 
characterized by larger R which can be interpreted as a larger relative 
star formation contribution to their MIR spectra. 
In particular the histogram points out a greater PAH detection rate in NLS1 than BLS1,
and, most of all that the majority of these detections correspond to the strongest 
circumnuclear SF activity (R$>$1). On the other hand R values for BLS1 are mostly upper limits, 
and the detections typically lie in the intermediate R rage (0.1$<$R$<$1).\\
This impression is substantiated by a 
formal censored statistics which we applied to the data. 
We used an improved version of the logrank test (Peto \& Peto~1972), 
which correctly treats left-censored data.
For the purpose of the analysis we divided the sample in two 
assuming in Fig.~\ref{fig3} a division line of FWHM(\hb)= 2000 \kms, 
and compare their properties with respect to R. 
This gave a negligible probability ($6\times10^{-5}$) that the two sub-samples 
are drawn from the same parent 
distribution. Moving the dividing line up or down by 100 \kms\ did not 
change the conclusion. This significant difference between NLS1 and BLS1 groups 
indicates that type-I AGNs with narrower broad emission lines 
reside in hosts containing more luminous SF regions. 

While the above is the main finding of the present paper, there are several 
potential effects that could bias this result and lower its significance. 
The ratio R could be small either due to dilution by the AGN continuum or 
or if the starburst is young (Rowan-Robinson \& Efstathiou 2009). 
The last possibility can be traced by relating the 9.7~$\mu$m silicate 
stenghts with the R values (similarly to Spoon et al. 2007). 
Howether, among our sample IRAS 11119+3257 and Mrk 231 are 
the only sources with strong silicate absorption, as shown in Fig. 10 and 
described in the Appendix for peculiar sources. 
Moreover, silicate emission is clearly detected in several sources 
(e.g. I Zw1, PG1211+143, PG0804+761 and others in both NLS1 and BLS1 samples), 
and well highlighted in the stacked spectra of Fig. 6. 
Thus, the possible contamination by young/obscured starbursts is mostly ruled out. 
The general issue of silicate emission and absorption features 
is the subject of a forthcoming paper including the present sample.\\
The two most important ones are aperture and luminosity effects. The first 
depends on the source distance and is the result of the fixed size 
\spitzer\ slit width that includes more host galaxy light in more distance 
sources. The IRS slit width is 3.6\arcsec\ for the spectral range used here, 
i.e 3.5~kpc at a typical distance of 200~Mpc.
 This can result in larger R (more PAH emission) at higher redshifts. 
The second potential bias is related to the well know increase in L(PAH) 
with AGN luminosity. The dependence of the two has been investigated in 
several samples (e.g. S06) and, while linear, the slope is not 1 which can 
introduce a trend of R with AGN luminosity. Our NLS1 and BLS1 samples where 
not chosen on the basis of their similar distances or AGN luminosity thus 
the above may affect the trend seen in Fig.~\ref{fig3}.
The rest of this section addresses the above potential biases by comparing 
the two samples in various ways that are relevant to those effects.

\subsection{Flux, luminosity and distance correlations}
\begin{figure*}[!h]
\begin{center}
% \vspace{-.5cm}
\includegraphics[scale=0.40]{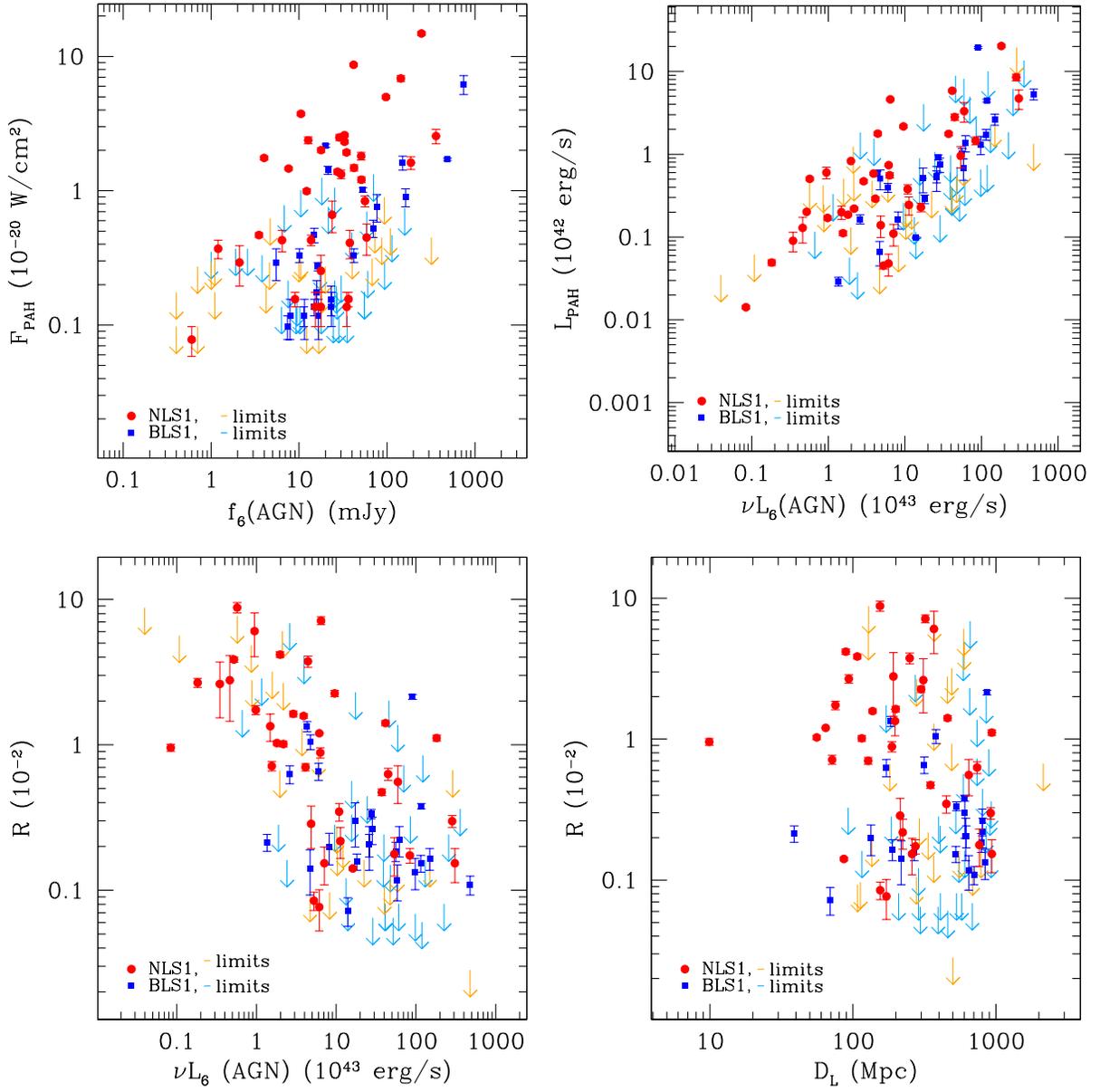}
\end{center}
% \linespread{1.0}
% \vspace{-.8cm}
\caption{Correlation plots for NLS1 (red) and BLS1 (blue). Symbols are as in Fig.~\ref{fig3}.
 Top left: 6.2~$\mu$m PAH flux as a function of 6~$\mu$m AGN flux. 
Top right: same as top left for luminosity.
Bottom left: The luminosity ratio R as a function L(6~$\mu$m).
 Bottom right: R as a function of source distances.} 
% \linespread{1.6}
\label{fig4}
\end{figure*}

We have constructed two-dimensional diagrams combining observational 
(measured flux and distances) and deduced (luminosities and R values) 
properties of the sources in our two sub-samples. These correlations are shown 
in Fig.~\ref{fig4}. The top left panel of the diagram examines the 
relation between the observed 6$\mu$m continuum and 6.2-PAH fluxes of our
sources. NLS1s and BLS1s seem to be broadly separated. A region defined by 
F(PAH)$\gtrsim 1\times 10^{-20}$~W~cm$^{-2}$ holds mostly NLS1. The two BLS1s
that are outside these limits are the ones with the strongest PAH, 
Mrk~231, and IRAS~13342+3932. Among these, Mrk~231 lies in the intermediate 
R values range of the histogram (Fig.~\ref{fig3}), while IRAS~13342+3932 (R$\sim2\times10^{-2}$)
represents the only BLS1 with SF activity comparable with what is observed in NLS1. 
Mrk~231 and IRAS~13342+3932 are also discussed in the Appendix for peculiar sources.
This separation strengthens the earlier suggestion of enhanced star formation 
activity, and the stronger PAH features in NLS1s, but may still be affected 
by aperture and luminosity effects. 

Possible physical trends are examined in the top right panel of 
Fig.~\ref{fig4}. This panel confirms that the earlier mentioned luminosity 
correlation between AGN continuum and star formation traced by PAH is also 
seen in our sample. Clearly L(PAH) increases with L(6~$\mu$m) for the entire 
sample. The slope of the correlation (which is not necessarily identical to 
the slope found in S06 who used a sample of higher luminosity sources) is 
$L(PAH) \propto L(6\mu m)^{0.7}$ i.e. sublinear.

The two bottom panels of the diagram examine the dependence of R on source 
luminosity and distance. The bottom left panel shows a clear decrease of R 
with L(6$\mu$m). This reflects the sublinear slope seen in the top right panel
meaning that L(6$\mu$m) increases more rapidly than L(PAH). The two 
sub-classes are still separated in this diagram, in the sense that at given 
AGN luminosity the NLS1 populate regions of larger luminosity ratio, with 
considerable scatter. Again this is suggesting enhanced star formation around
NLS1 AGN.\\
The bottom right panel of Fig.~\ref{fig4} is the one most relevant to the 
potential biases. Here R is plotted as a function of source distance which 
is translated to a larger physical aperture and increased host contribution.
The diagram shows no particular trends suggesting that host galaxy 
contribution is not an important factor in determining
R. The NLS1s-BLS1s separation in this plot is similar to the one seen in 
the other panels.

In conclusion, the NLS1 and BLS1 AGNs investigated here seem to represent 
two groups of type-I sources that differ in their star formation properties. 
This does not resemble an artifact of aperture or luminosity effects and seems to be 
an intrinsic property of the two sub-classes. The important parameter 
separating the two groups is their measured FWHM(\hb) (Fig.~\ref{fig3}). 
\begin{figure*}[!h]
\begin{center}
% \vspace{-.5cm}
\includegraphics[scale=0.50]{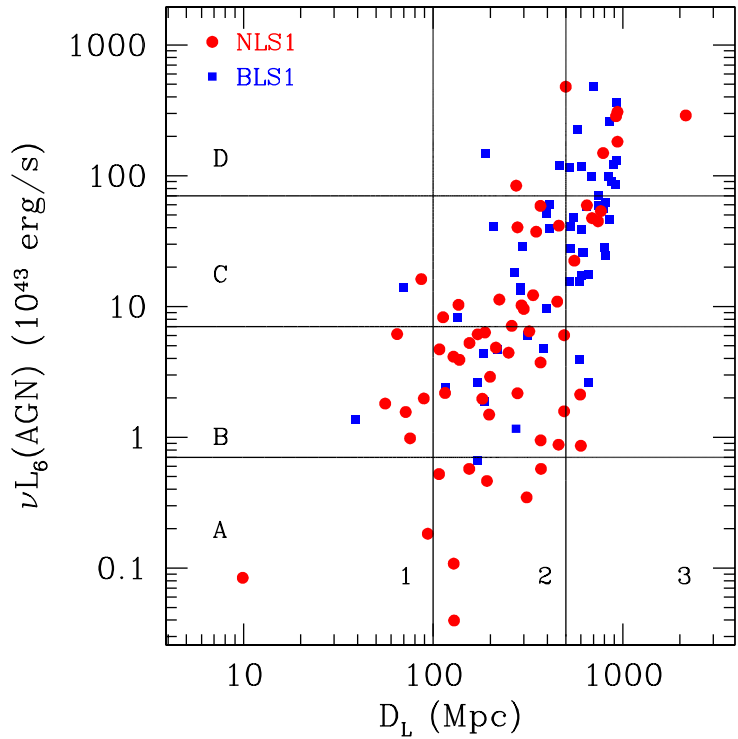}
\end{center}
% \linespread{1.0}
% \vspace{-.8cm}
\caption{Luminosity-Distance diagram. The 10 regions shown are defined considering the AGN luminosity and redshift orders of magnitude.} 
% \linespread{1.6}
\label{fig5}
\end{figure*}
To further investigate the potential influence of source distance and 
luminosity on the R vs. FWHM(\hb) correlation in a way fully including the PAH
nondetections, we have derived composite spectra of NLS1s and BLS1s in 
different luminosity-distance regimes. This was achieved by dividing the 
entire luminosity-distance space into ten regions defined by 
sampling the luminosity range in four equally spaced bins of the size of one order of magnitude, 
and the distance axis in three zones spacing the redshift range in bins of one order of magnitude each. 
This corresponds to distance thresholds of 100 and 500 Mpc and AGN luminosity limits of $7\times10^{42}$, $7\times10^{43}$ 
and $7\times10^{44}$ erg/s. The ten zones and the distribution of NLS1s and 
BLS1s in them are shown in Fig.~\ref{fig5}. We then obtained stacked 
spectra of all the NLS1s in each zone and compared them with stacked spectra 
of all the BLS1s in the same zone. 
The spectral combination is achieved by shifting all spectra to zero 
redshift and by normalizing individual spectra to the same 6$\mu$m AGN 
continuum flux density. The purpose is to inspect
visually the composite spectra in order to verify the previous conclusion
that R is indeed larger in NLS1s once the luminosity and distance effects have been removed.\\
The final number of sources for each zone and the composite spectra are shown in Fig.~\ref{fig6}.
%Zone 1 of this division 
%(D$_L<80$ Mpc, $\nu$L$_\nu$(6,AGN)$<3\times10^{43}$ erg/s) contains only 
%4 NLS1s and 4 BLS1s. Zone 4 ($80<$D$_L<200$ Mpc, 
%$3\times10^{43}<\nu$L$_\nu$(6,AGN)$<2\times10^{44}$ erg/s) has a low 
%%tatistics for BLS1s with only 2 sources. To increase the number of BLS1s 
%in this zone, we have included also the 2 sources within the same luminosity 
%range but at D$_L<80$ Mpc. The distances of these sources are close enough 
%to the limit of zone 4 and their inclusion in the analysis does not affect 
%the result. 
To compare quantitatively the star formation contributions of NLS1s and 
BLS1s in the same zone, we treated the composite spectra
in the same way used to define R for individual sources. We then define a 
variable describing the relative importance of star formation in NLS1 
compared to BLS1,
$F_{SB}$ by
\begin{equation}
 F_{SB}=\frac{R_{NLS1}(PAH, 6.2)}{R_{BLS1}(PAH, 6.2)},
\label{eq2}
\end{equation}
where $R_{NLS1}(PAH, 6.2)$ and $R_{BLS1}(PAH, 6.2)$ are the ratios of 
6.2~$\mu$m PAH fluxes to 6~$\mu$m AGN continuum (see Eq.~\ref{eq1}) for the NLS1 and BLS1 
average spectra, respectively.
$F_{SB}>1$ stands for a more intense SF activity in NLS1s 
compared to BLS1s. 
In Table~4 we list the measured values of R and F$_{SB}$ in all ten zones. 
We emphasize that in all the bins but two, an enhanced NLS1 star formation 
activity is found by at least a factor $\gtrsim~2$. 
Only one zone (A1) is an exception hosting no BLS1 and is discussed below.
\begin{figure*}[!h]
\begin{center}
% \vspace{-.5cm}
\includegraphics[scale=0.85]{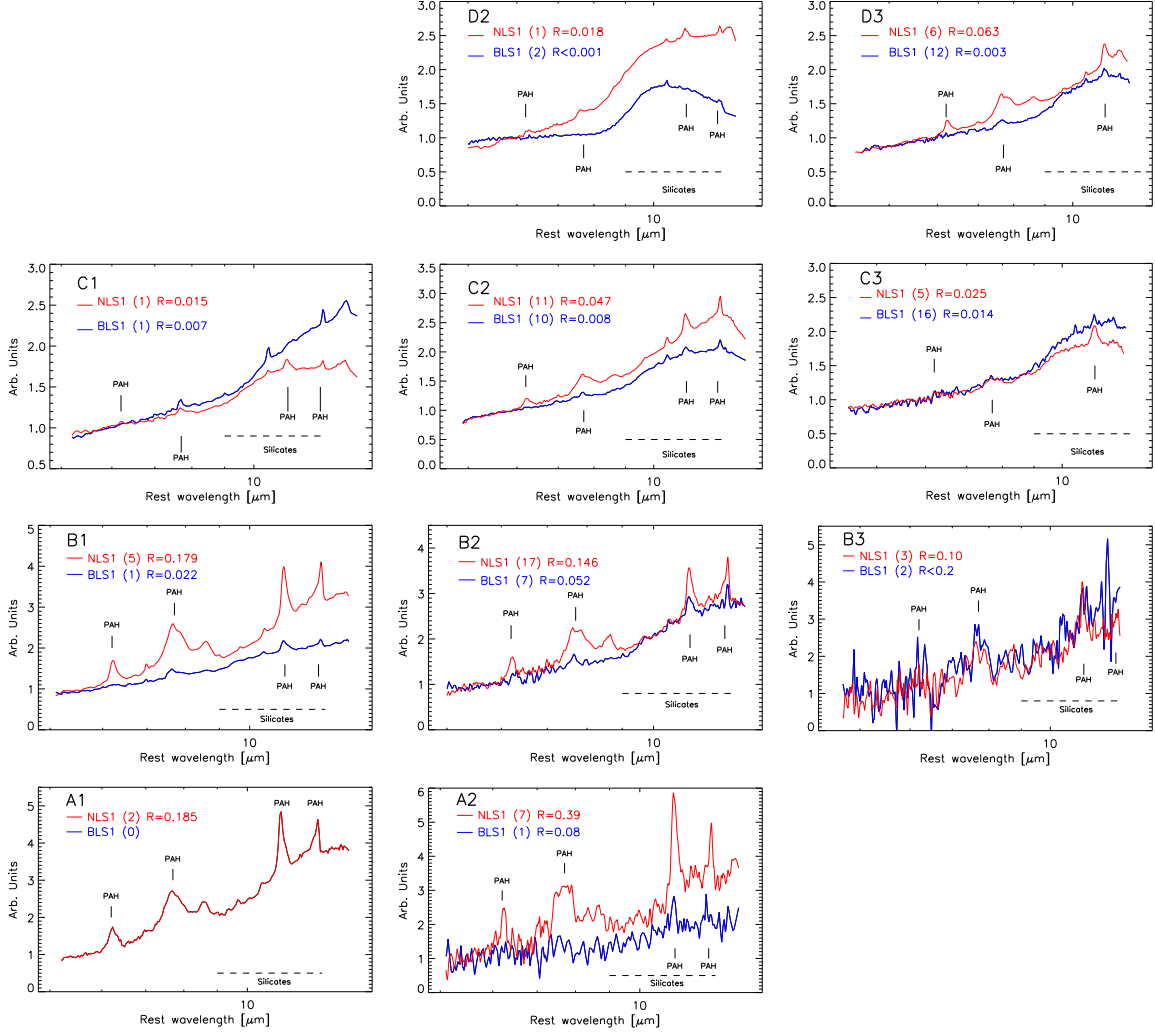}
\end{center}
% \linespread{1.0}
% \vspace{-.8cm}
\caption{Average NLS1 spectra (red) compared with averaged BLS1 spectra (blue). 
The panels are arranged and labelled like the regions defined in the luminosity-distance diagram (Fig.~\ref{fig5}). 
Each panels cites also the number of sources involved in the stacking procedure and the proper R values.} 
% \linespread{1.6}
\label{fig6}
\end{figure*}
A detailed account of the results of this analysis is in Appendix A (see Fig.~\ref{fig6} for a visual inspection). 
\begin{figure*}[!h]
\begin{center}
% \vspace{-.5cm}
\includegraphics[scale=0.75]{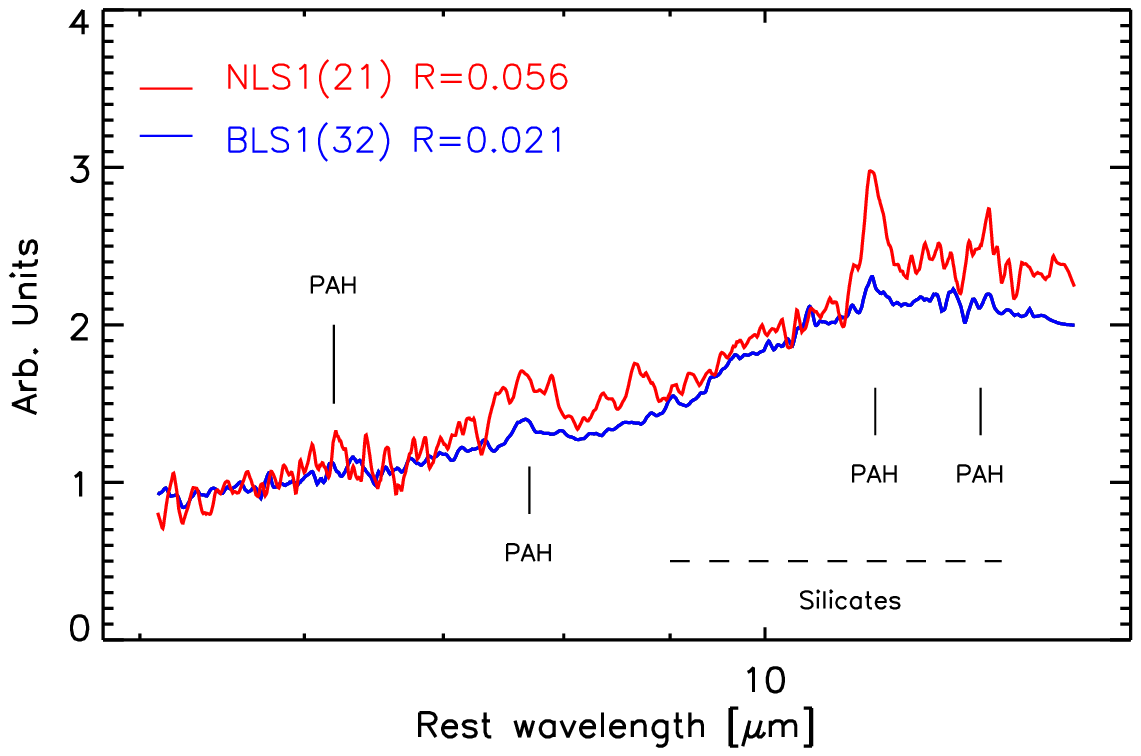}
\end{center}
% \linespread{1.0}
% \vspace{-.8cm}
\caption{Stacked spectra for NLS1 (red) and BLS1 (blue) characterized by no 6.2$\mu$m PAH detection in the individual spectrum.} 
% \linespread{1.6}
 \label{fig7}
\end{figure*}
To summarize, in almost all ten zones defined by distance 
and AGN luminosity, star forming activity in NLS1 seems to be stronger, by our 
definition based on the relative strength of the 6.2 PAH emission feature 
in NLS1s. The three zones (A1, C1, D2) with poor statistics in both NLS1 and BLS1 
require some caution. Nevertheless the small number of sources 
counts characterizing these zones is due to both physical and selection issues. 
Indeed, it is unexpected to find high luminosity sources (i.e. quasars) in 
the nearby Universe (z$<$0.02). 
We also note an intrinsic bias for sources entering a literature based archival sample: 
using the \hb\ width as the prime selecting criterion we may preferentially miss faint BLS1 galaxies, 
because broad \hb\ will be harder to identify against the host continuum. 
Finally the very faint and distant sources will not have entered the sample because 
of observational limitations even with a sensitive instrument like \textit{Spitzer}.
However, thanks to the sufficient statistics in most of the zones (8/10), 
the main result of stronger star formation activity in NLS1 is preserved and is clearly 
indicated in our R vs. FWHM(\hb) correlation (see Fig.\ref{fig3}). 
Also the combination of measured and deduced quantities
for NLS1s and BLS1s (Fig.~\ref{fig4}) robustly supports our finding.\\
As a further confirmation of our result, in Fig.~\ref{fig7} we have averaged the spectra of sources without
a detection of the 6.2~$\mu$m PAH feature. 
A clear detection of the 6.2~$\mu$m PAH is present in the stacked spectra of both NLS1 and BLS1, 
and a more intense star formation activity occurs in narrow line Seyferts 
with an NLS1/BLS1 enhance factor of F$_{SB}=2.6\pm0.5$.
\begin{figure*}[!h]
\begin{center}
% \vspace{-.5cm}
\includegraphics[scale=0.75]{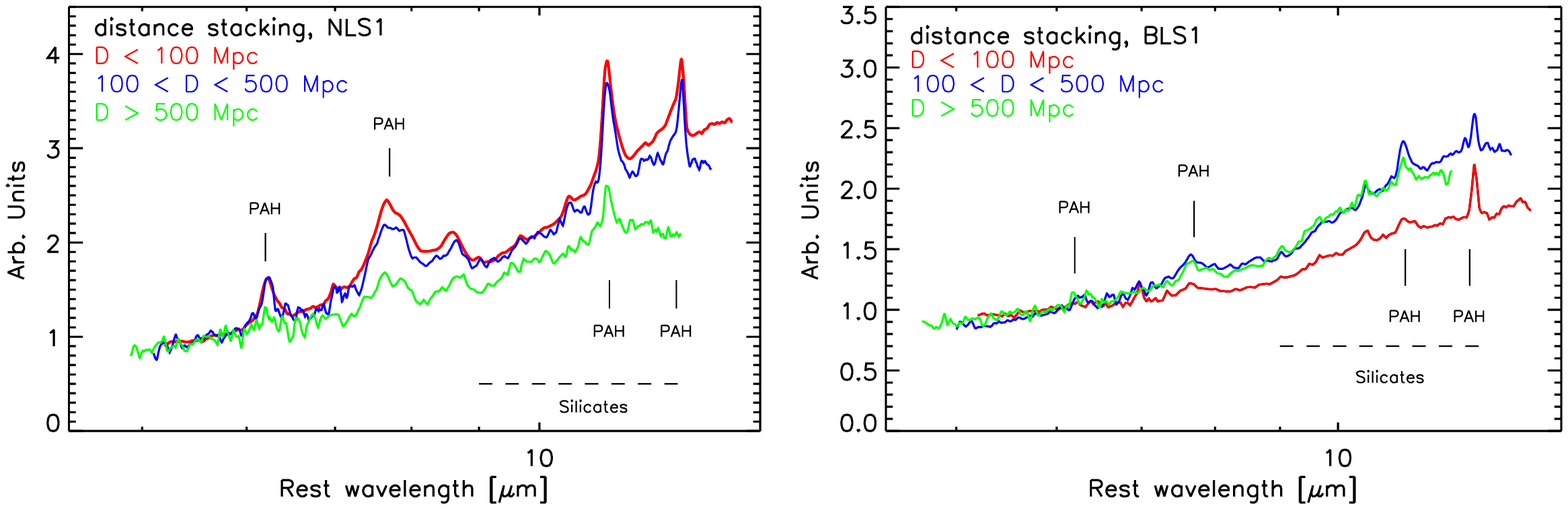}
\end{center}
% \linespread{1.0}
% \vspace{-.8cm}
\caption{Stacked spectra for NLS1 (\textit{left panel}) and BLS1 (\textit{right panel}). The data have been grouped in the three distance zones (1, 2, 3) of Fig.~\ref{fig5}.} 
% \linespread{1.6}
\label{fig8}
\end{figure*}
Concerning the possible luminosity and distance effects on the SF comparison factor, 
we do not find any peculiar trend. This can be inferred from both a visual inspection 
of different zones in Fig.~\ref{fig6}, and by comparing R and F$_{SB}$ values (Tab.~4) 
in terms of AGN luminosity and distance.\\ 
In particular, possible distance effect are crucial for the validity of our conclusion. 
%Indeed a contamination of the PAH features by the host galaxy would prevent us from the 
%investigation of AGN-SF physical connection (see \S6 for Details).
As a further check we perform a distance stack by collapsing the 
luminosity bins in the luminosity-distance plot (Fig.~\ref{fig5}) and grouping sources 
in the three distance zones. The mean distance-dependent spectra are shown in Fig.~\ref{fig8}. 
Significant distance effects would imply an increase of the 6.2 PAH intensity with increasing 
redshift. This effect is not observed in both NLS1 and BLS1 mean spectra. 
%On the contrary a faint PAH emission is revealed in the most distant (D$>500$~Mpc) NLS1s. The PAH faintness can be addressed to the intense quasar continuum dilution.
On the contrary, a relatively fainter PAH emission found for the most distant (D$>500$~Mpc) NLS1s. This can be plausibly ascribed to the higher AGN luminosity of these sources in combination with the sublinear increase of PAH luminosity with AGN luminosity (Fig.~\ref{fig4} top right). We note that, also for this grouping by distance only, NLS1 have stronger PAH.

\section{Discussion}
\begin{figure*}[!h]
\begin{center}
% \vspace{-.5cm}
\includegraphics[scale=0.4]{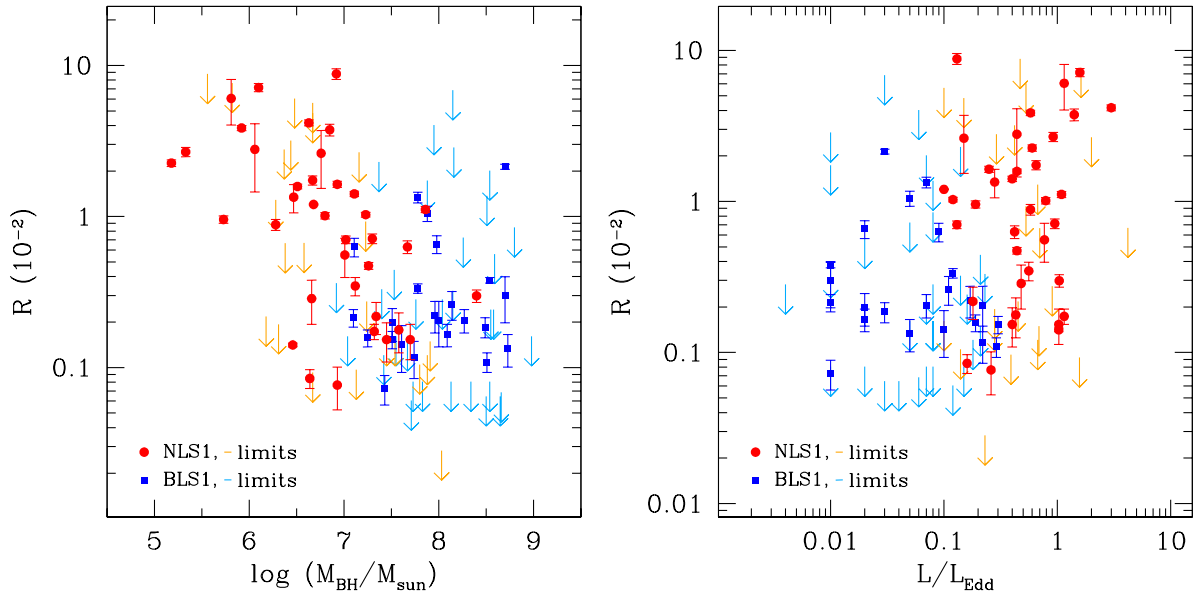}
\end{center}
% \linespread{1.0}
% \vspace{-.8cm}
\caption{PAH-to-AGN continuum ratio R as a function of the central SMBH parameters. Symbols are as in Fig.~\ref{fig3}.
\textit{Left panel}: R values versus central BH mass. \textit{Right panel:} R values versus Eddington luminosity ratio.
The BH masses and Eddington Rations are computed according to Eq.~\ref{eq3} and \ref{eq4} respectively (see text for details).} 
% \linespread{1.6}
\label{fig9}
\end{figure*}
In the previous section we have compared measured (flux and distance) and deduced (luminosity 
and R) parameters carefully taking into account luminosity and distance effects. 
A SF-AGN luminosity correlation has been seen and discussed in Netzer et al.~(2007) and Lutz et al.~(2008), and 
here we have confirmed this trend disentangling SF and AGN spectral components in the Spitzer short wavelength band.
We have demonstated an enhanced star formation activity in NLS1 galaxies compared to 
their broad line counterparts, and the enhancement factor (F$_{SB}$) 
does not appear to be affected by luminosity and distance effects.

In this section, we want to investigate a possible physical connection between the AGN fueling 
and star formation mechanisms. Such a connection has been proposed by several authors 
(i.e. Grupe \& Mathur~2004, Shemmer et al.~2004, Komossa \& Xu~2007 and references therein), 
but it has not been demonstrated by direct observational evidences. 
Here we compare the SF tracer with the AGN accretion parameters, that is, the 6.2 PAH luminosity -to- 
AGN continuum emission (R) is plotted as a function of the central black hole mass and Eddington ratio.\\
Black hole masses and accretion rates are computed by the single-epoch mass determination method based 
on the Kaspi et al.~(2005) using the Broad Line Region size vs 5100$\AA$ monochromatic luminosity (R$_{BLR}$-$\lambda$L$_\lambda$(5100)) relation. 
Thus, the BLR size is estimated from the measured $\lambda$L$_\lambda$(5100) (col.~5 of Tab.~\ref{tab1}), 
and the mass estimate follows from assuming Keplerian motion of the \hb\ emitting gas (FWHM(\hb\ ) are in col.~4 of Tab.~\ref{tab1}).
The reverberation mapping-based scaling relations adopted here are (see Netzer and Trakhtenbrot~2007 - NT07 for more details):
\begin{equation}
 M_{BH}=1.05\times10^8\left( \frac{L_{5100}}{10^{46}erg/s}\right)^{0.65} \left[ \frac{FWHM(H\beta)}{1000~km/s}\right]^2 M_\odot
\label{eq3}
\end{equation}
and
\begin{equation}
 L/L_{Edd}=\frac{fL_{5100}}{1.5\times10^{38}\left[ M_{BH}/M_\odot\right] }
\label{eq4}
\end{equation}
for BH mass and accretion rate respectively and where $L_{5100}$ stands for the 
5100$\AA$ monochromatic luminosity. In this equation $f$ is the bolometric 
correction factor. For low to intermediate luminosity AGN, this factor is between 7 and 12 
(Marconi et al.~2004, Netzer~2009) and is possibly luminosity dependent. 
Hence, bolometric luminosities were calculated from the monochromatic
luminosities ($L_{5100}$), by using the correction factors of Marconi et al.
(2004). These $M_{BH}$ (and thus $L/L_{Edd}$) values are considered to be accurate to
within a factor of $\sim2-3$ (see, e.g. Vestergaard \& Peterson 2006).\\
% %Here we do not consider the luminosity dependence and use $f=7$ for all sources according to NT07.\\
Figure~\ref{fig9} displays the bi-dimensional SF vs AGN properties plots. 
As seen in the left panel of this figure the star formation activity increases with decreasing BH masses, 
and moreover the region of low BH masses and highest 
R values (Log(M$_{BH}/$M$_\odot$)$<7$ and R$\gtrsim10^{-2}$)
is populated by NLS1s with the complete absence of BLS1s. 
Another interesting behavior is found also in the Fig.\ref{fig9} right panel, where the R 
value increases with the Eddington ratio, 
i.e. the AGN luminosity expressed in terms of the Eddington luminosity. 
Also in this case NLS1s are the only AGNs populating 
the region of the extreme measurements of both SF activity and Eddington ratio.\\
The upper limits in R make us refrain from expressing this trend in terms of a 
quantitative scaling relation between star formation activity and Eddington ratio.
Nevertheless, our plots demonstrate the supposed intimate correlation between 
the circumnuclear (within few central kpc) star formation and the AGN fueling.\\ 
The NLS1 galaxies are characterized by smaller BH mass, 
larger Eddington ratio and stronger SF activity 
compared to their broad line counterparts. Thus, we can safely conclude that AGN with the 
higher accretion efficiency are surrounded by more intense star formation 
than the less violently accreting SMBHs in the local Universe.\\
This SF-AGN correlation is possibly driven by a feedback process between the two. 
Chen et al.~(2009) have recently traced the star formation history of the circumnuclear regions 
in type~2 AGNs, and proposed a model in which supernova explosions play a key role in the transportation 
of gas to galactic centers. Other works (i.e Silk \& Rees~1998, Di Matteo et al.~2005) propose that 
AGN energetic outflows may affect the star formation in protogalaxies. 
While our work establishes a connection between star formation and accretion rate, 
we cannot draw definitive conclusions on its causal link to such feedback or feeding mechanisms.\\
%Even if our work addresses the SF-AGN connection, we can not draw definitive conclusion on these feedback. 
%Indeed the limited redshift range of our sample and the observable analyzed here can not determine which mechanism 
%between SF and BH growth drives the connection.\\
Shemmer et al.~(2004) demonstrated that NLS1s are also the local Universe 
representatives of the objects with high accretion rate (L/L$_{Edd}$) and high metallicity 
in the metallicity - accretion rate relationship for AGNs. Again this relation suggests an intimate 
connection between Starburst enriching the circumnuclear gas 
and AGN fueling, represented by the accretion rate. Because we demonstrate that the efficient SMBH growth in 
NLS1 is associated with the most active circumnuclear SF regions, we can only support the previous finding in an indirect way. 
Indeed the metallicity inferred by Shemmer and coauthors refer to the AGN broad line region, 
while the star forming activity investigated in our work happens in much larger regions.\\
From an evolutionary point of view, a high star formation efficiency is necessary 
for a rapid BH growth in high redshift systems (Kawakatu \& Wada 2009). 
Thus, assuming NLS1s as the early phase of BLS1s, our results would confirm the above model predictions. 
The local nature of our sources and the observable analyzed here can not directly confirm this evolutionary path, 
though our observations are an intriguing starting point for further high redshift investigations.
\section{Conclusions}
We have analyzed low resolution 5-15$\mu$m spectra of NLS1 and BLS1 galaxies obtained with the \textit{Spitzer} 
IRS spectrometer. 
Thanks to the high spatial resolution and sensitivity of the instrument we have determined AGN 6$\mu$m continuum and 6.2$\mu$m 
PAH luminosities.\\
From the comparison of observed and derived quantities (in particular the SF-to-AGN ratio) we find a 
more intense circumnuclear SF activity in NLS1s than in their broad line counterparts (Fig.~\ref{fig3}).
Possible luminosity and distance effect of this enhanced activity have been carefully taken into account by using 
both bi-dimensional diagrams (Fig.\ref{fig4}) and stacking procedures (Fig.~\ref{fig6}, \ref{fig8}). 
No trends affecting our main conclusion have been found.\\
Our paper shows yet another way to distinguish NLS1s from broader line
AGNs. This is another clear indication (on top of the extreme X-ray
properties) that emission line widths are indeed
related to fundamental physical properties, in the case under
study the properties of the host galaxies.\\
Finally, we have estimated the BH mass and accretion rate (Eq.~\ref{eq3}, \ref{eq4}) 
and plotted the SF-to-AGN ratio R as a function of these. 
Figure~\ref{fig9} shows that NLS1s hosting more violently accreting BH 
also harbor more intense circumnuclear star formation.
%
%\appendix

\section{Appendix A: stacked spectra}
The following is a detailed comparison of the NLS1 and BLS1 stacked spectra shown in Fig.~\ref{fig6}. 
\begin{description}
\item[Zone A1, D$_L<100$ Mpc, $\nu$L$_\nu$(6,AGN)$<7\times10^{42}$ erg/s.] 
These are nearby objects and the numbers are small with only 2 NLS1s 
and no BLS1s, thus preventing us from direct NLS1-BLS1 comparison for nearest 
and less luminous object. The absence of BLS1 sources is not surprising, 
indeed with such faint sources the detection of a broad \hb\ feature turns 
to be difficult due to the spread of the few revealed photons over a relatively 
large wavelength range. 
%Howether, given the low statistics of this bin, the final result remain valid. 
%
\item[Zone A2: $100<$D$_L<500$ Mpc, $\nu$L$_\nu$(6,AGN)$<7\times10^{42}$ erg/s.] 
Despite the low S/N of the spectra, here the result of the comparison is
 F$_{SB}=4.6\pm0.8$ i.e. stronger SF activity in NLS1s. The statistics is 
good for NLS1s while there is only one BLS1s. Again the faintness of sources 
and low S/N prevent from the detection of broad optical emitting lines. 
\item[Zone B1: D$_L<100$ Mpc, $7\times10^{42}<\nu$L$_\nu$(6,AGN)$<7\times10^{43}$ erg/s.]
Similarly to zone A2, the 
SF activity is enhanced in NLS1 (F$_{SB}=8.2\pm0.4$) even if we have to take in mind the low 
BLS1 statistics.
\item[Zone B2: $100<$D$_L<500$ Mpc, $7\times10^{42}<\nu$L$_\nu$(6,AGN)$<7\times10^{43}$ erg/s.]
As shown in Fig.~\ref{fig6}, the 6.2~$\mu$m PAH feature 
appears more prominent in the NLS1 mean spectrum than in the BLS1 one, 
and this is also confirmed by the relative importance of star formation 
F$_{SB}=2.8\pm0.2$.
\item[Zone B3: D$_L>500$ Mpc, $7\times10^{42}<\nu$L$_\nu$(6,AGN)$<7\times10^{43}$ erg/s.]
The low signal-to-noise of the data (lower-right panel of Fig.~\ref{fig6}) reflects the faintness of the sources.
Indeed we just compute a 3~$\sigma$ lower limit for the comparison parameter F$_{SB}>0.5$. 
Nevertheless, we note that the 6.2 PAH feature is detected in the NLS1 stacked spectrum. 

\item[Zone C1: D$_L<100$ Mpc, $7\times10^{43}<\nu$L$_\nu$(6,AGN)$<7\times10^{44}$ erg/s.]
A visual comparison of the data in upper-left panel of Fig.~\ref{fig6} does not allow to 
assign an enhanced SF activity to NLS1, but it is found from the comparison parameter 
F$_{SB}=2.2\pm0.2$. Despite the low statistics, the significance of the result 
is preserved by the high S/N and quality of the data. 
%Indeed, such spectra would determine the shape of the mean spectrum even with a slightly larger statistics.
%
\item[Zone C2: $100<$D$_L<500$ Mpc, $7\times10^{43}<\nu$L$_\nu$(6,AGN)$<7\times10^{44}$ erg/s.]
There are 10 NLS1s and 11 BLS1s in this intermediate distance zone.
The proper panel of Fig.~\ref{fig6} indicates a more intense 6.2~$\mu$m PAH 
feature in the mean NLS1 spectrum. The high statistics allows an accurate measurement of the SF comparison 
parameter: F$_{SB}=5.9\pm0.4$.
\item[Zone C3: D$_L>500$ Mpc, $7\times10^{43}<\nu$L$_\nu$(6,AGN)$<7\times10^{44}$ erg/s.]
The middle-right panel (Fig.~\ref{fig6}) shows the stacked spectra from 5 and 16 NLS1 and BLS1, respectively. 
All the sources involved in the stacking are bright PG quasars (Boroson \& Green~1992), and this nature 
reflects in the strong PAH dilution caused by the AGN continuum. 
PAH features are detected in both mean spectra, 
and a more intense SF activity is revealed in NLS1, (F$_{SB}=1.8\pm0.3$). 
\item[Zone D2: $100<$D$_L<500$ Mpc, $\nu$L$_\nu$(6,AGN)$>\times10^{44}$ erg/s.]
As in the case of zones A1 and C1, the statistics for both NLS1 and BLS1 is limited (respectively 1 and 2 sources). 
The 6.2~$\mu$m PAH feature is detected in the NLS1 spectrum, even if the quasar emission dilutes the line. 
On the contrary, in the BLS1 mean data the PAH feature is undetected. The corresponding lower limit 
for the comparison parameter is: F$_{SB}>20$ 
with high significance given the limiting R upper limit (R$<$0.001).
\item[Zone D3: D$_L>500$ Mpc, $\nu$L$_\nu$(6,AGN)$>7\times10^{44}$ erg/s.]
The most distant and luminous sources lie in this region. The statistics 
are good (6 NLS1s, 12 BLS1s), and the data have very good 
signal-to-noise. PHL~1092 is the most distant source in the entire sample 
(an isolate red point in D3 bin of Fig.~\ref{fig5}), and its SL observed 
spectrum does not cover the entire rest wavelength 
range, reaching only $\lambda_{rest}\sim11~\mu$m. As a consequence, the 
stacked NLS1 spectrum is distorted beyond 
$\lambda_{rest}\sim11~\mu$m. In the top-right panel of Fig.~\ref{fig6} we
plot the mean NLS1 spectrum obtained without PHL~1092 in order to 
show a more correct mean continuum shape for this zone.
This choice does not affect the shorter wavelength spectrum and the result
on relative importance of star formation (R values calculated with and without PHL~1092 
in the mean spectrum are consistent within the errors). 
The relative importance of star formation in zone D3 reaches the highest 
value (F$_{SB}=24\pm9$) supporting the main conclusion of this paper.
\end{description}

\section{Appendix B: peculiar sources}
\begin{figure*}[!h]
\begin{center}
% \vspace{-.5cm}
\includegraphics[scale=0.50]{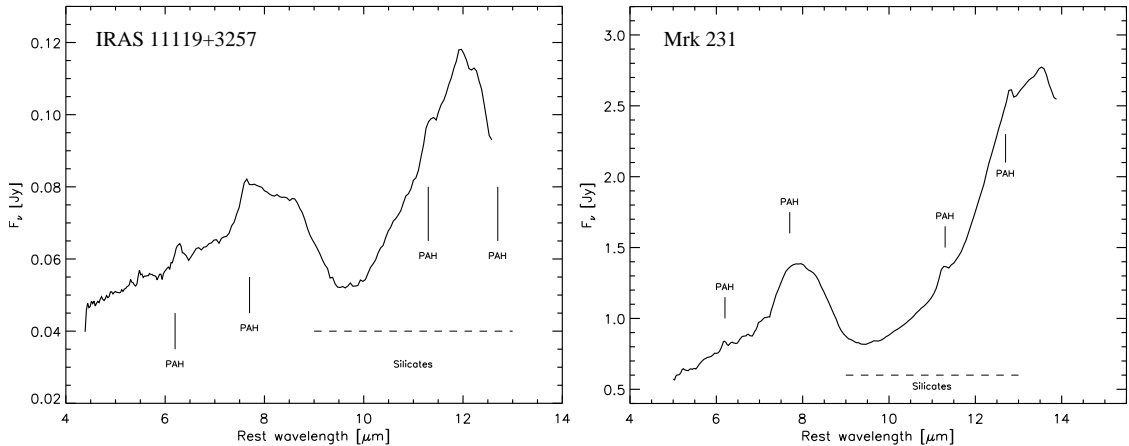}
\end{center}
% \linespread{1.0}
% \vspace{-.8cm}
\caption{Low Resolution mid-IR spectra of IRAS~11119+3257 and Mrk~231 showing the very strong
9.7~$\mu$m silicate absorption feature.} 
% \linespread{1.6}
\label{fig10}
\end{figure*}
- Mrk~335 (Fig.~\ref{fig2}) is a well characterized NLS1, 
but its mid-IR spectrum has no PAH features at all. 
Only a putative 7.7~$\mu$m PAH can be noticed, but it is distorted by a shallow 
Silicate absorption features centered at 9.7~$\mu$m, thus we can not confirm its detection. 
The shape of the Mrk~335 spectrum resembles that of quasars without PAH detections analyzed by S06. 
In that work the authors demonstrate that averaging 15 spectra similar to Mrk~335 
the 7.7 and 11.3 PAH peaks are detected on the top of the strong AGN continuum. 
The result leaves open the possibility that a modest circumnuclear SF activity is actually present in Mrk~335 
but the PAH features are strongly diluted by the AGN emission and lie under our detection capability.

- Among the archival samples there are three peculiar sources. Two of them 
(5-15~$\mu$m spectra are shown in Figure~\ref{fig10}), 
are the only objects showing a strong 9.7~$\mu$m silicate absorption:
a NLS1 (IRAS~11119+3257) and a BLS1 (Mrk~231), both characterized by ULIRG infrared emission.\\
Mrk~231 is a well known low-ionization broad absorption line quasar (lo-BAL QSO), 
with a broad line region optically detectable even if the nucleus is heavily obscured in X-rays (Gallagher et al.~2005). 
Its MIR spectrum shows not only the silicate absorption related to such an absorbed ULIRG, 
but also a luminous 6.2~$\mu$m PAH emission (L(6.2,PAH)$\sim3*10^{42}$erg/s) pointing out a significant starburst, 
even if not dominant (R=$0.17\times10^{-2}$). From the comparison of their low resolution spectra, 
IRAS~11119+3257 appears as the optical narrow line counterpart of Mrk~231. 
The only difference being the relative intensity of the Starburst, 
which in IRAS~11119+3257 is about twice the Mrk~231 values.\\
Even if it is not statistically significant, this coincidence is at least intriguing and may suggest a possible connection between NLS1 and lo-BAL quasars. Previous works support this suggestion: for example Brandt and Gallagher (2000) argued that the potential physical connection between NLS1s and lo-BAL QSOs is their high accretion rate close to the Eddington limit. Also Zheng et al. (2002) found common physical conditions (Fe II strength, [O III] weakness, strong outflows) in NLS1s, lo-BAL QSOs and IR QSOs and ascribed differences between them to different viewing angles and/or from different evolution phases.
\begin{figure*}[!h]
\begin{center}
% \vspace{-.5cm}
\includegraphics[scale=0.8]{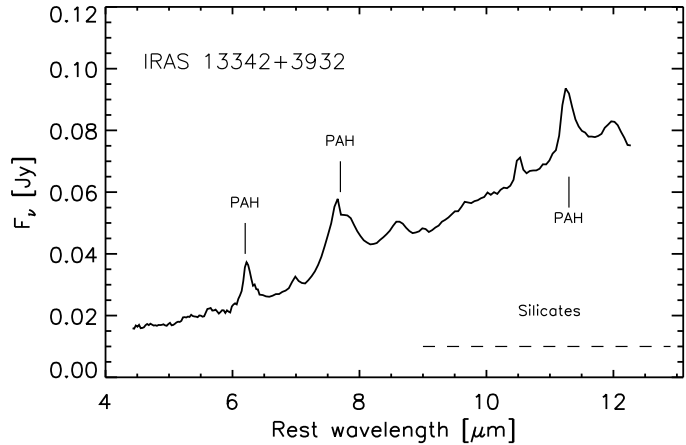}
\end{center}
% \linespread{1.0}
% \vspace{-.8cm}
\caption{Low Resolution mid-IR spectrum of IRAS~13342+3932, an optically classified BLS1 showing very strong
PAH emission features.} 
% \linespread{1.6}
\label{fig11}
\end{figure*}
\\
The third peculiar source in the archival sample is IRAS~13342+3932, another ULIRG galaxy, which is the only BLS1 
showing strong 6.2 PAH emission. 
Even if its optical spectrum is typical of a pure type 1 Seyfert, the mid-IR one (Fig.~\ref{fig11}) 
is Starburst dominated with an AGN infrared contribution $\sim15\%$ (Weedman \& Houck~2008). The fact that only 
one BLS1 is found with such a starburst dominated spectrum does not change our overall result about an enhanced star formation 
in NLS1 galaxies.
\section{Acknowledgments}
We thank the referee for helpful comments. 
We also thank Nico Hamaus for help with part of the initial data reduction.\\
Funding for this work at Tel Aviv University has been provided by the
Israel Science Foundation grant 364/07 and
the a DIP grant 15981.

\begin{table*}
\begin{tiny}
%\begin{center}
\centerline{\begin{tabular}{lcccc|lcccc}
\hline
\multicolumn{5}{c}{NLS1} &
\multicolumn{5}{c}{BLS1}\\
\hline
Source Name               & z        & D$_L$ & FWHM(H$_\beta$)  & Log $\lambda$L$_{5100}$ &
Source Name               & z        & D$_L$ & FWHM(H$_\beta$)  & Log $\lambda$L$_{5100}$\\
(1)                       & (2)      & (3)   & (4)             & (5) & 
(1)                       & (2)      & (3)   & (4)             & (5) \\
\hline
Ark 564                   & 0.0247   & 108   & $865^1$         & $44.12^2$ &
 Fairall 9                & 0.0470   & 208   & 6690$^{7}$      & 43.92$^{7}$\\
IRAS~13224-3809           & 0.0667   & 300   & $650^3$         & $44.81^2$ &
 IRAS 13342+3932          & 0.1790   & 866   & 7090$^7$        & 44.42$^7$\\
1H 0707-495               & 0.0411   & 181   & $1000^1$        & $43.48^4$ & 
 FBS 0732+396             & 0.1180   & 550   & $2370^7$        & $44.31^7$ \\ 
Ton S180                  & 0.0620   & 278   & $1090^6$        & $44.71^2$ & 
 J095504.55+170556.3      & 0.139    & 656   & $2460^7$        & $43.80^7$ \\
Mrk 1044                  & 0.0165   & 72    & $1310^1$        & $44.53^2$ & 
 J130842.24+021924.4      & 0.1400   & 661   & $5720^7$        & $43.86^7$ \\
Mrk 335                   & 0.0258   & 113   & $2081^7$        & $43.74^7$ & 
 J131305.68-021039.2      & 0.0837   & 381   & $4190^7$        & $43.86^7$  \\
Mrk 359                   & 0.0174   & 75    & $900^1$         & $43.73^2$ & 
 J141556.84+052029.5      & 0.1260   & 590   & $4130^7$        & $44.00^7$ \\
Mrk 896                   & 0.0264   & 115   & $1135^1$        & $43.95^2$ &
 IRAS F14463+3612         & 0.1130   & 525   & $2100^7$        & $44.08^7$ \\
Mrk 766                   & 0.0129   & 56    & $1100^6$        & $43.57^2$ &
 Mrk 836                  & 0.0388   & 171   & $7180^7$        & $43.14^7$\\
NGC 4051                  & 0.0023   & 10    & $1170^1$        & $42.26^2$ & 
 J152139.66+033729.2      & 0.1260   & 590   & $7310^7$        & $43.14^7$\\
PHL 1092                  & 0.3920   & 2150  & $1300^1$        & $44.45^4$ &
 J163631.29+420242.6      & 0.0610   & 273   & $7810^7$        & $43.47^7$ \\
PG 1244+026               & 0.0482   & 214   & $1270^7$        & $43.59^7$ &
 PG 0052+251              & 0.1550   & 739   & 5297$^7$        & 44.78$^7$ \\
RE J1034+393              & 0.0424   & 187   & $1050^7$        & $43.26^7$ &
 J171902.28+593715.9      & 0.1680   & 808   & $2250^7$        & $44.16^7$ \\
Mrk 1239                  & 0.0199   & 86    & $910^3$         & $43.72^2$ & 
 PG 1202+281              & 0.1650   & 792   & 6570$^7$        & 44.20$^7$ \\
Mrk 957                   & 0.0711   & 321   & $685^3$         & $43.55^4$ & 
 Mrk 1146                 & 0.0389   & 172   & $2700^7$        & $43.28^7$ \\
Mrk 507                   & 0.0559   & 250   & $960^3$         & $44.25^2$ &
 Mrk 590                  & 0.0265   & 116   & $7110^7$        & $43.87^7$\\
Mrk 291                   & 0.0352   & 155   & $2200^7$        & $43.25^7$ & 
 IRAS  07598+6508         & 0.1480   & 703   & $3150^{10}$     & $45.22^2$ \\
Mrk 493                   & 0.0313   & 137   & $1250^7$        & $43.37^7$ &
 VII Zw 244               & 0.1310   & 615   & $2950^7$        & $44.14^7$\\
PG 1448+273               & 0.0650   & 292   & $1060^7$        & $44.28^7$ &
 Mrk 106                  & 0.1230   & 575   & $4420^7$        & $44.51^7$ \\
Mrk 705                   & 0.0292   & 128   & $2270^7$        & $43.26^7$ &
 Mrk 1298                 & 0.0600   & 269   & $2380^7$        & $43.80^7$ \\
KUG 0301+002              & 0.0445   & 197   & $1460^7$        & $43.11^7$ &
 Ton 1542                 & 0.0640   & 287   & $3470^7$        & $43.90^7$ \\
J082912.67+500652.3       & 0.0434   & 192   & $1080^7$        & $42.88^7$ &
 NGC 4593                 & 0.0090   & 39    & $5040^7$        & $42.62^7$ \\
J024912.86-081525.6       & 0.0294   & 129   & $920^7$         & $44.33^7$ &
 Mrk 231                  & 0.0427   & 189   & $3130^7$        & $45.05^7$ \\
IRAS 04312+4008           & 0.0205   & 89    & $690^8$         & $44.36^8$ &
 Ton 730                  & 0.0870   & 397   & $3480^7$        & $43.93^7$ \\
Mrk 110                   & 0.0353   & 155   & $1900^7$        & $43.01^7$ &
 Mrk 279                  & 0.0305   & 134   & $5000^{14}$     & $43.06^4$ \\
J131305.80+012755.9       & 0.0294   & 128   & $2200^7$        & $43.84^7$ &
 PG 1415+451              & 0.1140   & 530   & $2970^7$        & $44.18^7$ \\
J094310.12+604559.1       & 0.0743   & 336   & $1100^7$        & $43.02^7$ & 
 PG 1519+226              & 0.1370   & 646   & $2380^7$        & $44.40^7$ \\
Mrk 142                   & 0.0450   & 199   & $1750^7$        & $43.57^7$ &
 PG 2209+184              & 0.0700   & 316   & $6690^7$        & $43.69^7$ \\
Mrk 42                    & 0.0246   & 107   & $940^7$         & $42.84^7$ &
 III Zw 2                 & 0.0893   & 408   & $4820^7$        & $44.41^7$ \\
J142748.28+050222.0       & 0.1060   & 490   & $1470^7$        & $44.26^7$ &
 Mrk 141                  & 0.0417   & 184   & $4910^7$        & $43.50^7$ \\
IC 3599                   & 0.0215   & 94    & $670^7$         & $42.40^7$ &
 PG 1149-110              & 0.0490   & 218   & $2920^7$        & $43.66^7$ \\
J170246.09+602818.9       & 0.0690   & 311   & $1980^7$        & $43.14^7$ &
 IC 4329A                 & 0.0160   & 69    & $5930^7$        & $43.17^7$ \\
J114008.71+030711.4       & 0.0811   & 368   & $550^7$         & $43.24^7$ &
 PG1416-129               & 0.1290   & 605   & 6885$^7$        & 44.12$^7$\\
J145047.19+033645.4       & 0.0670   & 302   & $1800^7$        & $42.99^7$ &
 PG 2304+042              & 0.0420   & 186   & $8390^7$        & $43.59^7$ \\
J124035.81-002919.4       & 0.0810   & 368   & $1000^7$        & $43.33^7$ &
 J151653.22+190048.2      & 0.1900   & 926   & $4840^{15}$     & $44.20^4$     \\
J125055.28-015556.8       & 0.0814   & 370   & $620^7$         & $43.24^7$ &
 PG 0804+761              & 0.1000   & 460   & $3085^7$        & $44.60^7$ \\
J135724.52+652505.8       & 0.1060   & 490   & $1240^7$        & $43.28^7$ &
 PG 0923+201              & 0.1900   & 926   & $6140^7$        & $45.06^7$  \\
J141234.68-003500.0       & 0.1270   & 595   & $1150^7$        & $43.44^7$ &
 PG 1411+442              & 0.0896   & 410   & 2850$^7$        & 44.31$^7$ \\
J144507.30+593649.9       & 0.1280   & 600   & $1960^7$        & $43.03^7$ &
 PG 1626+554              & 0.1320   & 620   & 4270$^7$        & 44.45$^7$  \\
IRAS 15462-0450           & 0.0998   & 459   & $1615^{10}$     & $43.96^2$ &
 PG 1048+382              & 0.1670   & 802   & 3550$^7$        & 44.49$^7$  \\
J172759.15+542147.0       & 0.0995   & 458   & $1380^7$        & $43.03^7$ &
 PG 1114+445              & 0.1440   & 682   & 5300$^7$        & $44.76^7$  \\
I Zw 1                    & 0.0611   & 274   & $1240^1$        & $44.63^9$ &
 PG 0844+349              & 0.0640   & 287   & 2300$^7$        & $43.97^7$ \\
PG 1211+143               & 0.0809   & 368   & $1860^5$        & $45.00^2$ &
 PG 1116+215              & 0.1770   & 856   & 3530$^7$        & $45.13^7$ \\
PKS 0558-504              & 0.1370   & 646   & $1250^1$        & $44.15^4$ &
 Mrk 1383                 & 0.0865   & 395   & 6620$^7$        & $44.45^7$ \\
IRAS11119+3257            & 0.1890   & 920   & $1980^{10}$     & $45.67^{10}$ &
 PG 1435-067              & 0.1260   & 590   & 3180$^7$        & $43.39^7$ \\
IRAS13349+2438            & 0.1080   & 500   & $2800^1$        & $44.64^6$ &
 Mrk 876                  & 0.1290   & 605   & 8810$^7$        & 44.67$^7$ \\
PG1404+226                & 0.0980   & 450   & $1390^7$        & $44.18^7$ &
 Mrk 877                  & 0.1140   & 530   & 5940$^7$        & 44.25$^7$ \\
Mrk 478                   & 0.0770   & 349   & $1450^7$        & $44.21^7$ &
 Mrk 304                  & 0.0658   & 296   & 5700$^7$        & 44.41$^7$ \\
PHL 1811                  & 0.1920   & 936   & $1500^{11}$     & $44.97^4$ &
 PG1322+659               & 0.1680   & 808   & 2850$^7$        & 44.51$^7$ \\
PG 0026+149               & 0.1450   & 687   & $1860^5$        & $44.96^5$ &
 PG 1151+117              & 0.1770   & 856   & 4770$^7$        & 44.70$^7$ \\
PG 1402+261               & 0.1640   & 786   & $2040^7$        & $44.30^7$ &
 PG1307+085               & 0.1550   & 739   & 4380$^7$        & 44.78$^7$ \\
PG 1115+047               & 0.1550   & 739   & $1860^7$        & $44.15^7$ &
 PG1309+355               & 0.1840   & 893   & 4910$^7$        & 45.08$^7$ \\
PG 1001+054               & 0.1600   & 765   & $1790^7$        & $44.55^7$ &
 IRAS 21219-1757          & 0.1130   & 525   & 2080$^{10}$     & 44.23$^{10}$ \\
PG 2130+099               & 0.0620   & 278   & $2800^7$        & $44.26^7$ &
 PKS 2349-01              & 0.1740   & 840   & 5480$^7$        & 44.82$^7$ \\
IRAS 03450+0055           & 0.0310   & 136   & $1310^8$        & $43.00^8$ &
 PG1012+008               & 0.1870   & 909   & 3220$^7$        & 44.63$^7$ \\
PG 1011-040               & 0.0580   & 259   & $1440^5$        & $44.30^5$ &
  &          &       &                 & \\
Mrk 734                   & 0.0502   & 223   & $2200^7$        & $43.88^7$ &
   &          &       &                 & \\
Mrk 486                   & 0.0389   & 172   & $1730^7$        & $43.59^7$ &
  &          &       &                 & \\
IRAS 20237-1547           & 0.1920   & 936   & $1570^{10}$     & $47.15^{10}$ &
                          &          &       &                 & \\
ESO 323-G077              & 0.0149   & 64    & $2100^{12}$     & $42.94^{13}$ &
                          &          &       &                 & \\
 PG 1552+085              & 0.1190   & 555   & $1670^7$        & $44.43^7$ &
                          &          &       &                 & \\
\hline
\end{tabular}}
%\end{center}
\caption{Overall properties of the complete sample. Columns: (1) source name; sources labeled with only the J2000 coordinates are from the SDSS catalog. The first 20 line in the NLS1 dedicated panel correspond to our original sample. (2) redshift. (3) Luminosity distance in Mpc for a H$_0 = 70$~km~s$^{-1}$Mpc$^{-1}$, $\Omega_m = 0.3$ and $\Omega_\Lambda = 0.7$ cosmology. (4)-(5) \hb~FWHM (km/s) and 5100~$\AA$ continuum luminosity (erg~s$^{-1}$) with related references. References: $^1$ Gallo~2006, $^2$ Hao et al.~2005, $^3$ Boller et al.~1996, $^4$ V{\'e}ron-Cetty \& V{\'e}ron~2006, $^5$ Boroson \& Green~1992, $^6$ Grupe et al.~2004, $^7$ this work: the electronically available optical spectra (i.e. SDSS sources and PG quasars) have been fitted consistently with NT07 \S2.2., $^8$ V{\'e}ron-Cetty \& V{\'e}ron~2001, $^9$ Zhou et al.~2006, $^{10}$ Zheng et al.~2002, $^{11}$ Whalen et al.~2006, $^{12}$ Schmid et al.~2003, $^{13}$ Mulchaey et al.~1996, $^{14}$ Boroson \& Meyers~1992, $^{15}$ Smith et al.~2003}
\label{tab1}
\end{tiny}
\end{table*}
\begin{table*}
 \begin{tiny}
  \centerline{\begin{tabular}{lcccc|lcccc}
 \hline
\multicolumn{5}{c}{NLS1} &
\multicolumn{5}{c}{BLS1}\\
 \hline
Source Name               & PID        & Date & Int. Time  & Ref &
Source Name               & PID        & Date & Int. time  & Ref\\
(1)                       & (2)        & (3)   & (4)         & (5) & 
(1)                       & (2)        & (3)   & (4)         & (5) \\
\hline
Ark 564                   & 20241      & 2005 Dec. 17 & 14x5 & 1 &
  Fairall 9                & 30572      & 2007 Jun. 16 & 14x2 & 1 \\
IRAS~13224-3809           & 20241      & 2005 Jul. 04 & 14x5 & 1 &
  IRAS 13342+3932          & 00105      & 2004 May. 13 & 60x2 & 4\\
1H 0707-495               & 20241      & 2006 Nov. 15 & 14x5 & 1 &
 FBS 0732+396             & 20741      & 2006 Apr. 23 & 14x4 & 1 \\
Ton S180                  & 20241      & 2005 Jul. 08 & 14x5 & 1 &
 J095504.55+170556.3      & 00049      & 2004 Apr.19  & 14x2 & 2 \\
Mrk 1044                  & 20241      & 2005 Aug. 04 & 14x5 & 1 &
 J130842.24+021924.4      & 20741      & 2006 Jan. 28 & 14x4 & 1 \\
Mrk 335                   & 20241      & 2005 Jul. 08 & 14x5 & 1 &
 J131305.68-021039.2      & 20741      & 2006 Jan. 31 & 14x4 & 1 \\
Mrk 359                   & 20241      & 2006 Jan. 15 & 14x5 & 1 &
 J141556.84+052029.5      & 20741      & 2066 Mar. 09 & 14x4 & 1 \\
Mrk 896                   & 20241      & 2005 Nov. 20 & 14x5 & 1 &
 IRAS F14463+3612         & 40991      & 2007 Jun. 20 & 60x1 & 1 \\
Mrk 766                   & 20241      & 2006 Jun. 22 & 14x5 & 1 &
 Mrk 836                  & 20741      & 2006 Jan. 27 & 14x4 & 1 \\
NGC 4051                  & 20241      & 2005 Dec. 11 & 14x5 & 1 &
 J152139.66+033729.2      & 20741      & 2005 Aug. 13 & 14x4 & 1 \\
PHL 1092                  & 20241      & 2006 Jan. 15 & 14x5 & 1 &
 J163631.29+420242.6      & 20741      & 2005 Aug. 15 & 14x4 & 1 \\
PG1244+026                & 20241      & 2005 Jul. 02 & 14x5 & 1 &
  PG1416-129               & 20142      & 2005 Aug. 15 & 60x2 & 3 \\
RE J1034+393              & 20241      & 2005 Dec. 16 & 14x5 & 1 &
 J171902.28+593715.9      & 40038      & 2007 Jun. 15 & 60x2 & 1 \\
Mrk 1239                  & 20241      & 2006 Dec. 19 & 14x5 & 1 &
  PG0052+251               & 00082      & 2004 Jan. 04 & 14x1 & 6\\ 
Mrk 957                   & 20241      & 2006 Jan. 01 & 14x5 & 1 &
 Mrk 1146                 & 20741      & 2006 Jan. 15 & 14x4 & 1 \\
Mrk 507                   & 20241      & 2005 Jul. 01 & 14x5 & 1 &
 Mrk 590                  & 00086      & 2006 Jan. 20 & 14x2 & 3 \\
Mrk 291                   & 20241      & 2005 Aug. 13 & 14x5 & 1 &
 IRAS  07598+6508         & 00105      & 2004 Feb. 29 & 14x3 & 4 \\
Mrk 493                   & 20241      & 2005 Aug. 15 & 14x5 & 1 &
 VII Zw 244               & 03187      & 2005 Nov. 14 & 60x2 & 5 \\
PG 1448+273               & 20241      & 2055 Jul. 01 & 14x5 & 1 &
 Mrk 106                  & 20741      & 2006 May. 20 & 14x4 & 1 \\
Mrk 705                   & 20241      & 2006 May. 28 & 14x5 & 1 &
 Mrk 1298                 & 03187      & 2005 Jul. 01 & 60x2 & 5 \\
%\hline
KUG 0301+002              & 20741      & 2006 Sep. 09 & 14x4 & 1 &
 Ton 1542                 & 03187      & 2005 Jul. 07 & 60x2 & 5 \\
J082912.67+500652.3       & 20741      & 2006 Apr. 24 & 14x4 & 1 &
 NGC 4593                 & 00086      & 2005 Jul. 02 & 14x2 & 3 \\
J024912.86-081525.6       & 30119      & 2007 Feb. 09 & 60x2 & 1 &
 Mrk 231                  & 00105      & 2004 Apr. 14 & 14x2 & 4 \\
IRAS 04312+4008           & 30715      & 2007 Mar. 24 & 14x4 & 1 &
 Ton 730                  & 20142      & 2006 Jan. 19 & 60x2 & 3 \\
Mrk 110                   & 20142      & 2006 May. 20 & 60x2 & 3 &
 Mrk 279                  & 00666      & 2003 Nov. 16 & 14x1 & 8 \\
J131305.80+012755.9       & 20741      & 2006 Jan. 28 & 14x4 & 1 &
 PG 1415+451              & 20142      & 2006 Jan. 18 & 60x2 & 3 \\
J094310.12+604559.1       & 20741      & 2006 May. 20 & 14x4 & 1 &
 PG 1519+226              & 20142      & 2005 Aug. 13 & 60x2 & 3 \\
Mrk 142                   & 20142      & 2006 May. 27 & 60x2 & 3 &
 PG 2209+184              & 20142      & 2006 Jun. 25 & 60x2 & 3 \\
Mrk 42                    & 30715      & 2007 Jun. 14 & 60x8 & 1 &
 III Zw 2                 & 00086      & 2005 Jul. 10 & 14x2 & 3 \\
J142748.28+050222.0       & 20741      & 2006 Jul. 24 & 14x4 & 1 &
 Mrk 141                  & 20741      & 2005 Nov. 15 & 14x4 & 1 \\
IC 3599                   & 30715      & 2006 Jun. 24 & 60x4 & 1 &
 PG 1149-110              & 20142      & 2005 Aug. 13 & 60x2 & 3 \\
J170246.09+602818.9       & 30119      & 2006 Aug. 07 & 60x2 & 1 &
 IC 4329A                 & 00086      & 2004 Jul. 13 & 14x2 & 3 \\
J114008.71+030711.4       & 30119      & 2007 Jun. 11 & 14x7 & 1 &
  PG1202+281               & 20142      & 2006 Jun. 22 & 60x2 & 3\\
J145047.19+033645.4       & 20741      & 2005 Aug. 15 & 14x4 & 1 &
 PG 2304+042              & 20142      & 2005 Dec. 17 & 60x2 & 3 \\
J124035.81-002919.4       & 20741      & 2005 Jul. 14 & 14x4 & 1 &
 J151653.22+190048.2      & 00049      & 2004 Apr. 01 & 14x1 & 2 \\
J125055.28-015556.8       & 30119      & 2006 Jul. 24 & 14x7 & 1 &
 PG 0804+761              & 00014      & 2004 Mae. 02 & 14x3 & 4 \\
J135724.52+652505.8       & 30119      & 2006 Dec. 23 & 60x2 & 1 &
 PG 0923+201              & 03187      & 2005 Nov. 12 & 60x2 & 5 \\
J141234.68-003500.0       & 30119      & 2006 Jul. 24 & 14x7 & 1 &
 PG 1411+442              & 03187      & 2005 Jan. 11 & 60x2 & 5 \\
J144507.30+593649.9       & 30119      & 2006 Dec. 23 & 60x2 & 1 &
 PG 1626+554              & 03187      & 2005 Jan. 08 & 60x5 & 5 \\
IRAS 15462-0450           & 00105      & 2004 Mar. 03 & 60x1 & 4 &
 PG 1048+382              & 20142      & 2006 Jan. 31 & 60x2 & 3 \\
J172759.15+542147.0       & 30119      & 2006 Aug. 04 & 60x2 & 1 &
 PG 1114+445              & 20142      & 2006 May. 07 & 60x2 & 3 \\
I Zw 1                    & 00014      & 2004 Jan. 07 & 14x2 & 4 &
 PG 0844+349              & 03187      & 2005 Nov. 14 & 60x2 & 5 \\
PG 1211+143               & 00014      & 2004 Jan. 08 & 14x3 & 4 &
 PG 1116+215              & 00082      & 2004 May. 14 & 6x1 & 6 \\
PKS 0558-504              & 00082      & 2004 Mar. 27 & 14x1 & 1 &
 Mrk 1383                 & 03187      & 2005 Feb. 13 & 60x2 & 5 \\
IRAS 11119+3257           & 00105      & 2004 May. 13 & 60x1 & 4 &
 PG 1435-067              & 03187      & 2007 Jul. 12 & 60x2 & 5 \\
IRAS 13349+2438           & 00061      & 2005 Jun. 07 & 14x5 & 7&
 Mrk 876                  & 03187      & 2004 Nov. 15 & 60x2 & 5 \\
PG1404+226                & 20142      & 2006 Mar. 05 & 60x2 & 3 &
 Mrk 877                  & 03187      & 2005 Mar. 15 & 60x3 & 5 \\
Mrk 478                   & 03187      & 2005 Jan. 14 & 60x2 & 5 &
 Mrk 304                  & 03187      & 2004 Nov. 15 & 60x3 & 5 \\
PHL 1811                  & 30426      & 2006 Nov. 12 & 14x2 & 1 &
 PG 1322+659              & 20142      & 2006 Jun. 21 & 60x2 & 3 \\
PG 0026+149               & 03187      & 2005 Aug. 11 & 240x4 & 5 &
 PG 1151+117              & 00082      & 2003 Dec. 18 & 6x1 & 6 \\
PG 1402+261               & 00082      & 2004 Jun. 08 & 14x1 & 6 &
 PG1307+085               & 00082      & 2006 Jan. 19 & 6x1 & 3 \\
PG 1115+047               & 20142      & 2006 May. 27 & 60x2 & 3 &
 PG1309+355               & 00082      & 2004 May. 14 & 6x1 & 6 \\
PG 1001+054               & 03187      & 2005 May. 23 & 240x2 & 5 &
 IRAS 21219-1757          & 03187      & 2004 Nov. 16 & 60x2 & 5 \\
PG 2130+099               & 00014      & 2004 Jun. 06 & 14x3 & 4 &
 PKS 2349-01              & 03187      & 2004 Dec. 13 & 60x3 & 5 \\
IRAS 03450+0055           & 00082      & 2004 Aug. 31 & 6x1 & 7 &
 PG1012+008               & 20142      & 2006 dec. 19 & 14x7 & 3 \\
PG 1011-040               & 20142      & 2006 Jun. 23 & 60x2 & 3 &
                          &            &              &      &    \\
Mrk 734                   & 00014      & 2004 May. 13 & 14x3 & 4 &
                           &            &              &      &   \\
Mrk 486                   & 03421      & 2005 Jan. 06 & 60x3 & 9 &
                          &            &              &      &    \\
IRAS 20237-1547           & 00105      & 2004 Apr. 19 & 60x2 & 4 &
                          &            &              &      &    \\
ESO 323-G077              & 30745      & 2006 Aug. 06 & 14x2 & 1 &
                          &            &              &      &   \\
 PG 1552+085              & 20142      & 2005 Aug. 13 & 60x2 & 3 &
                          &            &              &      &   \\
\hline
\end{tabular}
}
 \end{tiny}
\caption{Observation log. Columns: (1) Source Name. (2) Proposal identification code. (3) Observation Date. (4) Exposure Time (ramp duration in sec per number of cycles). (5) First publication of the IRS/Spitzer low resolution data: 1. This work, 2. Shy et al.~2007, 3. Shy et al.~2006, 4. Hao et al.~2005, 5. Schweitzer et al.~2006, 6. Maiolino et al.~2007, 7. Thompson et al.~2009, 8. Dasyra et al.~2008, 9. Cao et al.~2008.}
\label{tab2}
\end{table*}

\begin{table*}
\begin{tiny}
%\begin{center}
\centerline{\begin{tabular}{lccccc|lccccc}
\hline
\multicolumn{6}{c}{NLS1} &
\multicolumn{6}{c}{BLS1}\\
\hline
Source Name              &  f$_\nu$ (AGN)  &  F$_{PAH}$   & $\nu$L$_\nu$ (AGN)  & L$_{PAH}$ & R   &
Source Name              &  f$_\nu$ (AGN)  &  F$_{PAH}$   & $\nu$L$_\nu$ (AGN)  & L$_{PAH}$ & R  \\
%                         &     mJy     & $10^{-20}$ W/cm$^2$ & $10^{43}$erg/s  & $10^{42}$erg/s & $10^{-2}$ &
%                         &     mJy     & $10^{-20}$ W/cm$^2$ & $10^{43}$erg/s  & $10^{42}$erg/s & $10^{-2}$ \\
(1)                      & (2)             & (3)           & (4)                & (5)       & (6) & 
(1)                      & (2)             & (3)           & (4)                & (5)       & (6) \\
\hline
Ark 564                  & 67.6            &  $<$0.31       & 4.70              & $<$0.043 & $<$0.09 &
   Fairall 9               & 159             & $<$0.74        & 41.2              & $<$0.39  & $<$0.09 \\
IRAS~13224-3809          & 17.8            & $2.01\pm0.08$  & 9.58              & 2.17     & 2.26 &
 IRAS 13342+3932          & 20.2            & $2.16\pm0.06$  & 90.6              & 19.4     & 2.15\\
1H 0707-495              & 10.0            &  $<$0.33       & 1.97              & $<$0.13  & $<$0.66 & 
 FBS 0732+396            & 26.5            & $<$0.20        & 47.9              & $<$0.71  & $<$0.15 \\
Ton S180                 & 4.7             & $<$0.62        & 2.17              & $<$0.58  & $<$2.66 & 
 J095504.55+170556.3     & 6.8             & $<$0.78        & 17.5              & $<$4.0   & $<$2 \\
Mrk 1044                 & 50.9            &  $1.81\pm0.12$ & 1.56              & 0.11     & 0.71 & 
 J130842.24+021924.4     & 1.0             & $<$0.35        & 2.62              & $<$1.8   & $<$7.0 \\
Mrk 335                  & 108.9           & $<$0.53        & 8.27              & $<$0.080 & $<$0.10 & 
 J131305.68-021039.2     & 5.5             & $0.29\pm0.08$  & 4.77              & 0.51     & 1.1 \\
Mrk 359                  & 28.8            &  $2.50\pm0.14$ & 0.98              & 0.17     & 1.34 & 
 J141556.84+052029.5     & 1.9             & $<$0.37        & 3.95              & 0.15     & $<$3.9 \\
Mrk 896                  & 27.4            & $1.38\pm0.06$  & 2.18              & 0.22     & 1.01 &
 IRAS F14463+3612        & 9.5             & $<$0.16        & 15.6              & $<$0.51  & $<$0.33 \\
Mrk 766                  & 97.4            & $4.99\pm0.16$  & 1.81              & 0.19     & 1.03 &
 Mrk 836                 & 3.8             & $<$0.33        & 0.66              & $<$0.12  & $<$1.8 \\
NGC 4051                 & 144             & $6.9\pm0.3$    & 0.084             & 0.008    &  0.96 &
 J152139.66+033729.2     & 7.5             & $<$0.21        & 15.6              & $<$0.90  & $<$0.6 \\
PHL 1092                 & 10.5            & $<$0.35        & 289               & $<$19    & $<$0.67 &
 J163631.29+420242.6     & 2.6             & $<$0.37        & 1.16              & $<$0.33  & $<$2.8 \\
PG1244+026               & 17.7            & $0.25\pm0.08$  &  48.5             & 0.14     & 0.29 &
PG1416-129              & 7.9             & $0.12\pm0.04$  & 17.3              & 0.51     & 0.30\\
RE J1034+393             & 30.1            & $1.3\pm0.1$    & 6.33              & 0.56     & 0.88 &
J171902.28+593715.9      & 6.1             & $<$0.12        & 23.8              & $<$0.9   & $<$0.4 \\
Mrk 1239                 & 361             & $2.5\pm0.3$    & 16.2              & 0.23     & 0.14 & 
PG0052+251              & 21.5            & $<$0.76        & 70.2              & $<$5.0   & $<$0.71\\
Mrk 957                  & 10.5            & $3.74\pm0.12$  & 6.46              & 4.62     & 7.14 & 
 Mrk 1146                & 14.9            & $0.47\pm0.06$  & 2.62              & 0.17     & $<$0.63 \\
Mrk 507                  & 12.7            & $2.38\pm0.14$  & 4.73              & 1.77     & 3.75 &
 Mrk 590                 & 30.2            & $<$0.23        & 2.42              & $<$0.04  & $<$0.2\\
Mrk 291                  & 4.0             & $1.76\pm0.06$  & 0.57              & 0.50     & 8.79 & 
IRAS  07598+6508         & 162.5           & $0.90\pm0.14$  & 480               & 5.30     & 0.11 \\
Mrk 493                  & 32.9            & $2.59\pm0.08$  & 3.71              & 0.58     & 1.58 &
VII Zw 244               & 11.5            & $0.12\pm0.04$  & 26.0              & 0.53     & 0.20 \\ 
PG 1448+273              & 19.9            & $<$0.27        & 10.2              & $<$0.28  & $<$0.27 &
Mrk 106                  & 114             & $<$0.47        & 225               & $<$1.9   & $<$0.08 \\
Mrk 705                  & 42.2            & $1.48\pm0.08$  & 4.13              & 0.29     & 0.70 &
Mrk 1298                 & 42.1            & $0.33\pm0.04 $ & 18.2              & 0.29     & 0.16 \\ 
%\hline
KUG 0301+002             & 6.4             & $0.43\pm0.08$  & 1.49              & 0.20     & 1.34 &
 Ton 1542                & 28.3            & $<$0.12        & 14.0              & $<$0.12  & $<$0.08 \\
J082912.67+500652.3      & 2.1             & $0.3\pm0.1$    & 0.46              & 0.13     & 2.79 &
 NGC 4593                & 150.8           & $1.6\pm0.2$    & 1.36              & 0.03     & 0.21 \\
J024912.86-081525.6      & 0.4             & $<$0.18        & 0.04              & 0.03     & $<$8.8 &
 Mrk 231                 & 746             & $6\pm1$        & 159               & 2.65     & 0.17 \\ 
IRAS 04312+4008          & 4.17            & $8.7\pm0.2$    & 1.98              & 0.83     & 4.17  &
 Ton 730                 & 10.2            & $<$0.14        & 9.61              & 0.26     & 0.27\\ 
Mrk 110                  & 36.5            & $0.16\pm0.02$  & 90.6              &  0.04    & 0.005 &
 Mrk 279                 & 76.7            & $0.8\pm0.2$    & 8.20              & 0.16     & 0.2 \\ 
J131305.80+012755.9      & 1.1             &  $<$0.31       & 0.11              & $<$0.06  & $<$5.7 &
 PG 1415+451             & 16.4            & $0.27\pm0.02$  & 27.5              & 0.92     & 0.33 \\
J094310.12+604559.1      & 18.0            & $<$0.19        & 12.2              & $<$0.26  & $<$0.22 & 
 PG 1519+226             & 23.4            & $0.14\pm0.04$  & 58.4              & 0.68     & 0.12 \\ 
Mrk 142                  & 12.2            & $0.99\pm0.04$  & 2.90              & 0.47     & 1.63 &
 PG 2209+184             & 10.1            & $0.33\pm0.04$  & 6.02              & 0.40     & 0.66 \\
Mrk 42                   & 7.6             & $1.46\pm0.04$  &  0.52             & 0.20     & 3.85 &
 III Zw 2                & 39.8            & $<$0.49        & 39.6              & $<$0.47  & $<$0.25 \\
J142748.28+050222.0      & 4.2             & $<$0.2         & 6.03              & $<$0.56  & $<$ 0.93 &
 Mrk 141                 & 21.3            & $1.4\pm0.1$    & 43.3              & 0.58     & 1.34 \\
IC 3599                  & 3.5             & $0.47\pm0.02$  & 0.18              & 0.05     & 2.68 &
 PG 1149-110             & 16.6            & $0.12\pm0.04$  & 4.70              & 0.07     & 0.14 \\
J170246.09+602818.9      & 0.6             & $0.08\pm0.02$  &  0.35             & 0.09     & 2.60 &
 IC 4329A                & 486             & $1.72\pm0.03$  & 14.0              & 0.10     & 0.07 \\
J114008.71+030711.4      & 1.2             & $0.37\pm0.06$  &  0.97             & 0.60     & 6.18 &
  PG1202+281              & 14.9            & $0.14\pm0.02$  & 55.8              & 1.02     & 0.18 \\
J145047.19+033645.4      & $<$0.6          & $<$0.21        & $<$0.33           & $<$0.23  & - &
PG 2304+042              & 9.2             & $<$0.14        & 1.89              & $<$0.06  & $<$0.3 \\
J124035.81-002919.4      & 4.6             & $<$0.29        &  3.73             & $<$0.47  & $<$1.3 &
 J151653.22+190048.2     & 70.2            & $<$1.3         & 360               & $<$14    & $<$0.4     \\
J125055.28-015556.8      & 0.7             & $<$0.27        & 0.57              & $<$0.45  & $<$7.8 &
 PG 0804+761             & 94.0            & $<$0.29        & 119               & $<$0.74  & $<$0.06 \\
J135724.52+652505.8      & 1.1             & $<$0.18        & 15.8              & $<$0.50  & $<$3.2 &
 PG 0923+201             & 25.5            & $<$0.21        & 131               & $<$2.20  & $<$0.2  \\
J141234.68-003500.0      & 1.0             & $<$0.29        & 2.12              & $<$1.24  & $<$5.9 &
 PG 1411+442             & 60.4            & $<$0.25        & 60.6              & $<$0.51  & $<$0.08 \\
J144507.30+593649.9      & 0.4             & $<$0.10        & 0.86              & $<$0.42  & $<$4.9 &
 PG 1626+554             & 11.3            & $0.12\pm0.02$  & 26.0              & 0.54     & 0.21  \\
IRAS 15462-0450          & 32.9            & $2.32\pm0.08$  &  41.5             & 5.86     & 1.41 & 
 PG 1048+382             & 7.4             & $0.10\pm0.02$  & 28.5              & 0.75     & 0.26  \\
J172759.15+542147.0      & 0.7             & $<$0.10        & 0.88              & $<$0.25  & $<$2.8 &
  PG 1114+445            & 35.2            & $<$0.12        & 97.9              & $<$0.65  & $<$0.07  \\
I Zw 1                   & 187             & $1.62\pm0.18$  & 83.9              & 1.45     & 0.17 & 
 PG 0844+349             & 27.0            & $<$0.18        & 13.3              & $<$0.17  & $<$0.13 \\
PG 1211+143              & 72.8            & $<$0.55        & 58.8              & $<$0.88  & $<$0.15 &
 PG 1116+215             & 59.0            & $<$0.70        & 258               & $<$6.2   & $<$0.24 \\
PKS 0558-504             & 23.8            & $0.66\pm0.18$  & 59.4              & 3.31     & 0.56 &
 Mrk 1383                & 55.6            & 0.18           & 51.7              & $<$0.33  & $<$0.06 \\
IRAS 11119+3257          & 56.2            & $0.84\pm0.08$  & 285               & 8.50     & 0.30 &
  PG 1435-067            & 17.8            & $<$0.14        & 37.0              & $<$0.57  & $<$0.15 \\
IRAS 13349+2438          & 321             & $<$0.45        & 479               & $<$1.34  & $<$0.03 & 
  Mrk 876                & 53.4            & $1.01\pm0.04$  & 117               & 4.45     & 0.38 \\
PG 1404+226              & 9.0             & $0.16\pm0.02$  & 10.9              & 0.38     & 0.35 &
 Mrk 877                 & 24.5            & $<$0.12        & 41.1              & $<$0.39  & 0.10  \\
Mrk 478                  & 51.2            & $1.21\pm0.06$  & 37.3              & 1.76     & 0.47 &
 Mrk 304                 & 55.1            & $<$0.18        & 28.8              & $<$0.18  & $<$0.06 \\
PHL 1811                 & 58.7            & $0.45\pm0.12$  &  308              & 4.71     & 0.15 &
 PG1322+659              & 15.9            & $0.18\pm0.04$  & 62.0              & 1.37     & 0.22  \\
PG 0026+149              & 16.8            & $<$0.10        & 47.4              & $<$0.55  & $<$0.12 &
 PG 1151+117             & 10.5            & $<$1.0         & 46.0              & $<$8.9   & $<$1.9  \\
PG 1402+261              & 40.2            & $<$0.35        & 149               & $<$2.6   & $<$0.18 &
 PG1307+085              & 18.1            & $<$1.3         & 59.1              & $<$8.2   & $<$1.4 \\
PG 1115+047              & 13.7            & $0.43\pm0.04$  &  44.8             & 2.81     & 0.63 &
PG1309+355               & 25.5            & $<$1.1         & 122               & $<$10    & $<$0.8 \\
PG 1001+054              & 15.3            & $0.14\pm0.04$  & 53.6              & 0.96     & 0.18 &
 IRAS 21219-1757         & 69.6            & $0.53\pm0.08$  & 115               & 1.73     & 0.15  \\
PG 2130+099              & 87.2            & $<$0.45        &  40.3             & $<$0.42  & $<$0.10 &
 PKS 2349-01             & 23.2            & $0.16\pm0.04$  & 97.8              & 1.32     & 0.13  \\
IRAS 03450+0055          & 93.5            & $<$0.90        &  10.3             & $<$0.20  & $<$0.19 &
 PG1012+008              & 17.3            & $<$0.30        & 85.5              & $<$2.9   & $<$0.34  \\
PG 1011-040              & 17.7            & $0.14\pm0.04$  &  7.11             & 0.11     & 0.16 &
  &                  &                &                   &          & \\
Mrk 734                  & 37.8            & $0.4\pm0.1$    & 11.3              & 0.24     & 0.22 &
 &                  &                &                   &          &   \\
Mrk 486                  & 34.8            & $0.14\pm0.04$  &  6.13             & 0.05     & 0.08 &
 &                  &                &                   &          &  \\
IRAS 20237-1547          & 34.7            & $1.93\pm0.08$  & 182               & 20.3     & 1.11 &
 &                  &                &                   &          & \\
ESO 323-G077             & 247             & $14.8\pm0.4$   & 6.15              & 0.74     & 1.2  &
 &                  &                &                   &          & \\
PG 1552+085             & 12.2             & $<$0.10        & 22.4              & $<$0.36  & $<$0.16 &
                        &                  &                &                   &          & \\
\hline
\end{tabular}}
\caption{
Mid-IR spectral parameters. Columns: (1) source name; sources labeled only with J2000 coordinates are from the SDSS catalog. (2) AGN flux density at 6~$\mu$m in units of mJy. Sources with a 6~$\mu$m upper limit are nondetections at a 3~$\sigma$ level and are not included in the bidimensional plots. (3) 6.2~$\mu$m PAH flux in $10^{-20}$W/cm$^2$. (4) AGN $6~\mu$m luminosity in $10^{43}$erg/s. (5) PAH luminosity in $10^{42}$erg/s. (6) SB/AGN luminosity ratio in units of $10^{-2}$.
}
\end{tiny}
\label{tab3}
\end{table*}
\begin{table*}[!h]
 \begin{center}
 \begin{tabular}{lccc}
\hline
\hline
 Zone & R~(NLS1) & R~(BLS1) & F$_{SB}$ \\
\hline
A1 & 0.185$\pm0.003$ & - & - \\
A2 & 0.387$\pm0.012$ & 0.085$\pm0.012$ & 4.6$\pm0.8$ \\
B1 & 0.179$\pm0.002$ & 0.022$\pm0.001$ & 8.2$\pm0.4$ \\
B2 & $0.146\pm0.004$ & 0.052$\pm0.003$ & 2.8$\pm0.2$ \\
B3 & 0.10$\pm0.02$  & $<0.2$ & $>0.5$ \\
C1 & 0.015$\pm0.001$ & 0.007$\pm0.001$ & 2.2$\pm0.2$ \\
C2 & 0.047$\pm0.01$ & 0.008$\pm0.004$ & 5.9$\pm0.4$  \\
C3 & 0.025$\pm0.001$ & 0.014$\pm0.002$ & 1.8$\pm0.3$ \\
D2 & 0.018$\pm0.001$  & $<0.001$ & $>20$ \\
D3 & 0.063$\pm0.001$ & 0.003$\pm0.001$ & 24$\pm9$ \\
\hline
\end{tabular}
\end{center}
\begin{footnotesize}
\caption{Star formation activity parameters for the ten stacking zones in Fig.~\ref{fig6}: R and F$_{SB}$ are shown with propagated errors. 
The error on F$_{SB}$ are statistical, the dispersion of the data is not taken into account.}
\end{footnotesize}
\label{tab4}
\end{table*}
%
%
%\begin{figure*}[!h]
%\begin{center}
%% \vspace{-.5cm}
%\includegraphics[scale=0.50]{histogram.ps}
%\end{center}
% \linespread{1.0}
% \vspace{-.8cm}
%\caption{histogream} 
% \linespread{1.6}
%\label{fg:h_pah}
%\label{fig:hist}
%\end{figure*}
%
%%%%%%%%%%%%%%%%%%%%%%%%%%%%%%%%%%%%%%%%%%%%%%%%%%%%%%%%%%%%%%
\newpage
%

%%%%%%%%%%%%%%%%%%%%%%%%%%%%%%%%%%%%%%%%%%%%%%%%%%%%%%%%%%%%% 

\end{document}